\documentclass{aa}
\usepackage{graphicx}

\begin{document}
\title{Simulating the recent star formation history in the halo of NGC~5128
\thanks{Based 
on observations collected at the European Southern Observatory, Paranal,
Chile, within the Observing Programme 63.N-0229}} 

\author{M. Rejkuba\inst{1}
	\and L. Greggio\inst{2}
	\and M. Zoccali\inst{1,3}
}

\offprints{M. Rejkuba}

\institute{European Southern Observatory, Karl-Schwarzschild-Strasse
           2, D-85748 Garching, Germany\\
           E-mail: mrejkuba@eso.org, mzoccali@eso.org 
	\and INAF, Osservatorio Astronomico di Padova, 
		Vicolo dell'Osservatorio 5, 35122 Padova, Italy \\
	   E-mail: greggio@pd.astro.it
        \and Departamento de Astronom\'{\i}a y Astrof\'{\i}sica, Pontificia Universidad
	     Cat\'olica de Chile, Casilla 306, Santiago 22, Chile; and Princeton 
             University Observatory, Peyton Hall, Princeton, NJ08544-1001 \\
	}

\date{15 August 2003 / 6 November 2003}
\titlerunning{Simulated SFH in Cen A halo}

\abstract{%
Simulated color-magnitude diagrams are used to investigate the recent
star formation history in NGC~5128.  The comparison of the simulations
with the observed $UV$ color-magnitude diagram for a field in the
north-eastern shell, where recent star formation is present,
constrains the initial mass function (IMF) and the duration of the
star formation episode.  The star formation burst is still on-going or
at most has stopped some 2 Myr ago. The look-back time on the main
sequence is set by the incompleteness of the $U$-band observations at
$V \sim 26$, and is of the order of $50$~Myr. The post main sequence
phases have a look-back time up to 100~Myr, setting the maximum
observable time for the star formation in this field. The comparison
of the observed and simulated luminosity functions and the number of
post main sequence vs. total number of stars favours models with
active star formation in the last 100~Myr. 
%The absence of gaps on the
%main sequence indicates that probably no major fluctuations in the
%star formation rate have occurred over the sampled time interval.  
The data are best fitted by an IMF with Salpeter slope ($\alpha=2.35$),
and are also consistent with slopes in the range of $2\la \alpha \la
2.6$. They exclude steeper IMF slopes.  The mean star formation rate
for a Salpeter IMF in the range of masses $0.6\le {\rm M} \le
100$~M$_\odot$ within the last 100~Myr is $1.6 \times 10^{-4} {\rm
M}_\odot\, {\rm yr}^{-1}\, {\rm kpc}^{-2}$.
}

\maketitle

%
%-------------------------------------------------------------------
%

\section{Introduction}

The two main traditional models of elliptical galaxy formation, 
the ``monolithic collapse'' (Partridge \& Peebles \cite{partridge&peebles67}) 
and the ``hierarchical assembly'' model (Toomre \cite{toomre77}) 
assume very different mechanisms and epochs of spheroid assembly. 
Correspondingly,
they also predict different mean ages and metallicities for the stars that
build these galaxies. Until the advent of the HST and 10m class ground-based
telescopes the studies of the stellar populations in the 
elliptical galaxies were
restricted to integrated light photometry and spectroscopy. 
In the last few years, however, the nearest E/S0 galaxies have been
resolved into stars (e.g. Soria et al. \cite{soria+96}, 
Davidge \& van den Bergh \cite{davidge+vandenbergh01}, 
Schulte-Ladbeck et al. \cite{schulte-ladbeck+03}). The study of the 
star formation histories of the resolved stellar populations observed in these 
galaxies is an important complementary approach to the studies of the 
high redshift Universe, aiming to address the questions 
of elliptical galaxy formation and evolution.

\subsection{Recent Star Formation in the halo of NGC 5128}

At the distance of 3.8~Mpc (Soria et al.~\cite{soria+96}, 
Rejkuba \cite{rejkuba04}) NGC 5128 (=Centaurus A) is the dominant
galaxy of the nearby Centaurus group and one of the closest giant elliptical
galaxies to us. It is also one of the nearest AGNs and 
radio galaxies. It presents a number of peculiarities: the central 
dust lane with HII regions and OB associations at its edge, 
and the stellar shells which have presumably been formed in a 
recent accretion of a smaller companion galaxy 
(see review by Israel \cite{israel98}). Moreover, young blue supergiants 
have been identified associated with the optical
filaments in the north-eastern halo (Graham \& Price 
\cite{graham&price81}, Graham \cite{graham98}). 

The young stars and OB associations in the halo of NGC~5128 are roughly 
aligned with the ionized gas and 
the radio jet (Morganti et al.~\cite{morganti+99}, Mould et al.~\cite{mould+00},
Fassett \& Graham~\cite{fg00}, Rejkuba et al.\ \cite{rejkuba+01, rejkuba+02})
over several kiloparsecs.
The preferred mechanism for this recent star formation in the
halo at $>14$ kpc away from the nucleus is through the direct interaction of 
the radio jet with the gas. The necessary amounts of neutral gas for
this star formation have been detected in nearby fields through HI 
(Schiminovich et al.~\cite{schiminovich+94}) and CO
molecular transitions (Charmandaris et al.~\cite{charmandaris+00}). 
Charmandaris et al.\ suggested that the molecular clouds in the distant halo
are associated with stellar shells and are thus remains of the last accretion
episode in which NGC~5128 has merged with a gas rich companion galaxy.
While the gas could have just ``hung around'' in the galaxy halo, the presence
of the radio jet triggered the star formation in highly collimated areas. 
The investigation of the star formation history of the young stars and the 
determination of the age spread in the young stellar population
can constrain the formation of the radio jet or the 
time of the last accretion event.
The metallicity of the gas of the accreted
galaxy should be reflected in the newly formed stars. 

The comparison of the optical color-magnitude diagrams (CMDs) with stellar
evolutionary isochrones and young LMC clusters has indicated that these young
stars can be as young as 10-15 Myr old (Mould et al.~\cite{mould+00}, 
Fassett \& Graham \cite{fg00}, 
Rejkuba et al.\ \cite{rejkuba+01}). However, except for a comparison
with one synthetic CMD (Mould et al.~\cite{mould+00}), no detailed studies
of the star formation history in the halo of NGC~5128 were made.

\subsection{The Method}

Containing stars that were born during the whole
life-time of the galaxy, CMDs represent snapshots of their star formation
histories. Even when limited to the bright stars, the CMDs contain useful 
age and metallicity indicators: the colors of the red giant branch (RGB) stars
are very sensitive to metallicity; the presence of bright asymptotic giant 
branch (AGB) stars signals 
an intermediate-age population; the colors and magnitudes of the bright 
main sequence (MS) stars and blue supergiants depend on the most recent 
star formation history, as well as on the metallicity.

The most effective way to analyse the distribution of stars in the
CMD and to quantify the star formation 
history of a galaxy is through the comparison of the observed CMD with
synthetic CMDs computed using stellar evolution models. This technique
is based on the simulations of a large number of CMDs constructed 
assuming different star formation rates
(SFR($t$)), chemical enrichment laws ($Z(t)$) and initial mass functions
(IMF). The photometric errors and the incompleteness of the data are also
simulated. Each simulated CMD then depends on all the main parameters which
determine the distribution of stars in a CMD. In this way it is best
accounted for the evolutionary behaviour of stars, small number of stars, and
spread due to observational effects. 
The synthetic CMDs are then compared to the observed ones in order to
determine the acceptable range of parameters and the best fitted models.  
This technique is currently used by several groups to investigate the star 
formation history (SFH) in dwarf galaxies 
%was first described by the Bologna group 
(Ferraro et al.~\cite{ferraro+89}, Tosi et al.~\cite{tosi+91},
%), and 
%later applied, with some variations in the 
%details, to investigate different LMC
%stellar fields and clusters as well as dwarf galaxies by a number of groups 
%(
Bertelli et al.~\cite{bertelli+92}, 
Tolstoy \cite{tolstoy96}, Gallart et al.~\cite{gallart+96a,gallart+96b},
Dolphin \cite{dolphin97}, Greggio et al.~\cite{greggio+98},
Hurley-Keller, Mateo \& Nemec~\cite{hurley-keller+98},
Hern\'andez, Valls-Gabaud \& Gilmore \cite{hvg98}, 
Cole et al.~\cite{cole+99},  Olsen \cite{olsen99},
Schulte-Ladbeck et al.~\cite{schulte-ladbeck+00},
Harris \& Zaritsky \cite{harris&zaritsky}). 

Here we apply this technique to investigate the recent star formation history
of the halo of NGC~5128. The observations were taken
with VLT using FORS1 and ISAAC in the U, V and Ks bands. Their full
description as well as data reduction and the resulting CMDs 
were presented in Rejkuba et al.~(\cite{rejkuba+01}). 

This paper starts with the description of the synthetic CMD simulator,
including a detailed description of the algorithms used. 
In the next section the 
young stellar population in the north-eastern halo field is simulated.
We test two different metallicities and a range of star formation
histories. The initial mass function is also investigated. 

%
%----------------------------------------------------------------------
%
\section{Description of the synthetic CMD simulator}

Our synthetic CMD simulator is developed based on the code used by 
Greggio et al.\ (\cite{greggio+98}). 
The construction of a synthetic CMD goes through the following steps:
\begin{enumerate}

\item Random extraction of mass according to the initial mass
function (IMF), age according to a chosen star formation rate (SFR) law,
and metallicity according to the chemical enrichment law ($Z(t)$).

\item Positioning of the extracted star as a function of mass, age, and 
metallicity on the theoretical HR
diagram ($\log({\rm L}/{\rm L}_\odot)$, $\log({\rm T}_{\rm eff})$) 
interpolating within stellar evolutionary tracks.

\item Transformation of the theoretical position of the star in the HR
diagram ($\log({\rm L}/{\rm L}_\odot)$, $\log({\rm T}_{\rm eff})$) 
into the observational CMD assigning the
bolometric corrections through interpolation within conversion tables.

\item Application of observational uncertainties, photometric errors and
incompleteness, according to the results of artificial star experiments on the
real data.

\item Iteration until a certain criterion 
(e.g. the total number
of stars generated, the total mass extracted, or the total number of stars
in a certain region of the simulated CMD)
is satisfied. 
\end{enumerate} 

The algorithms used for steps 1 and 2 are relatively straightforward, and used
in a similar way by most groups. Hence we describe them in the Appendix. 
In the following sections we address steps 3 and 4 in more detail as the 
final results depend critically on them.

%
%----------------------------------------------------------------------
%

\subsection{Bolometric corrections}
\label{bolcorr_sect}

\begin{figure*}
\centering
\includegraphics[width=8.5cm]{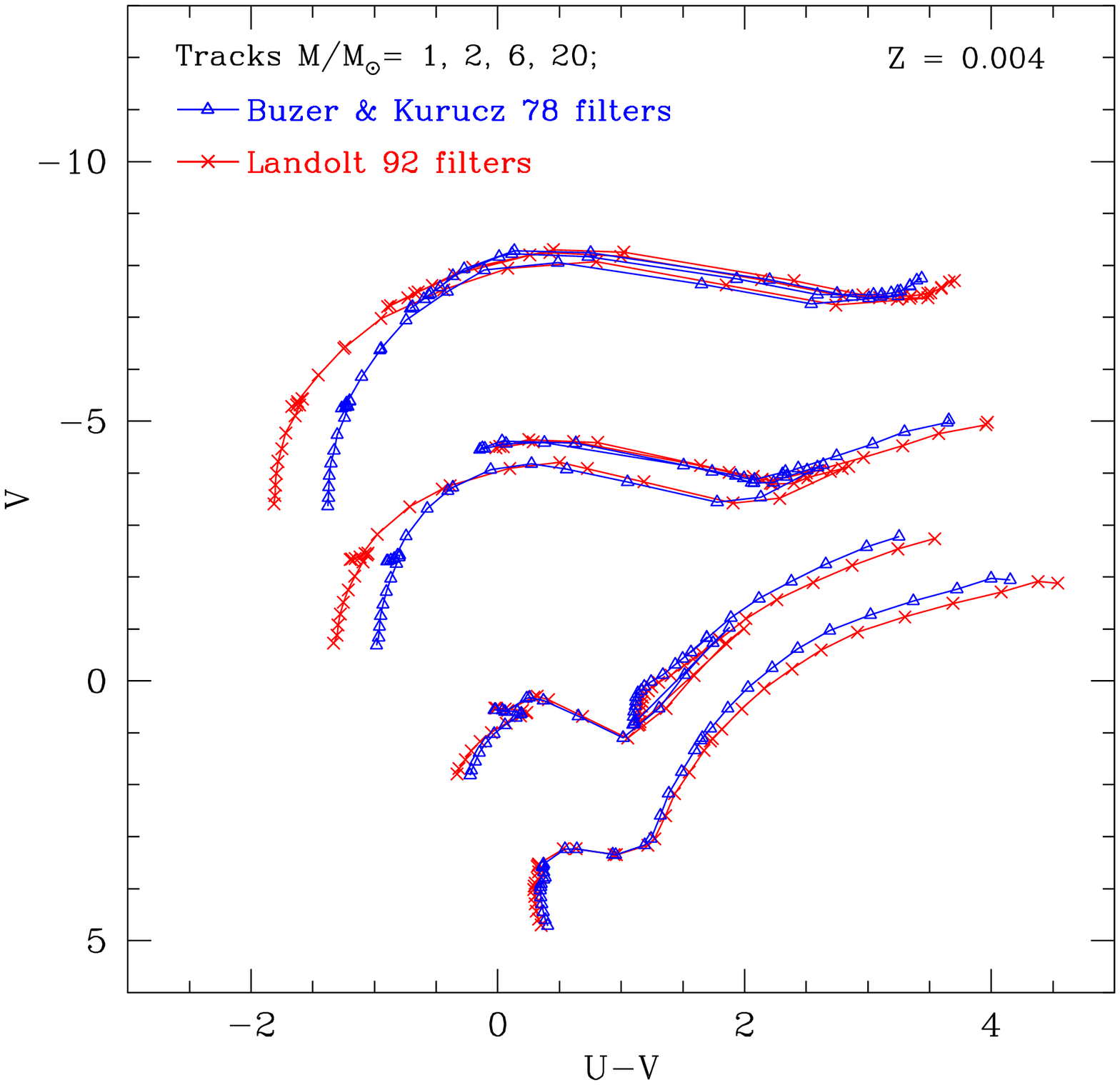}
\includegraphics[width=8.5cm]{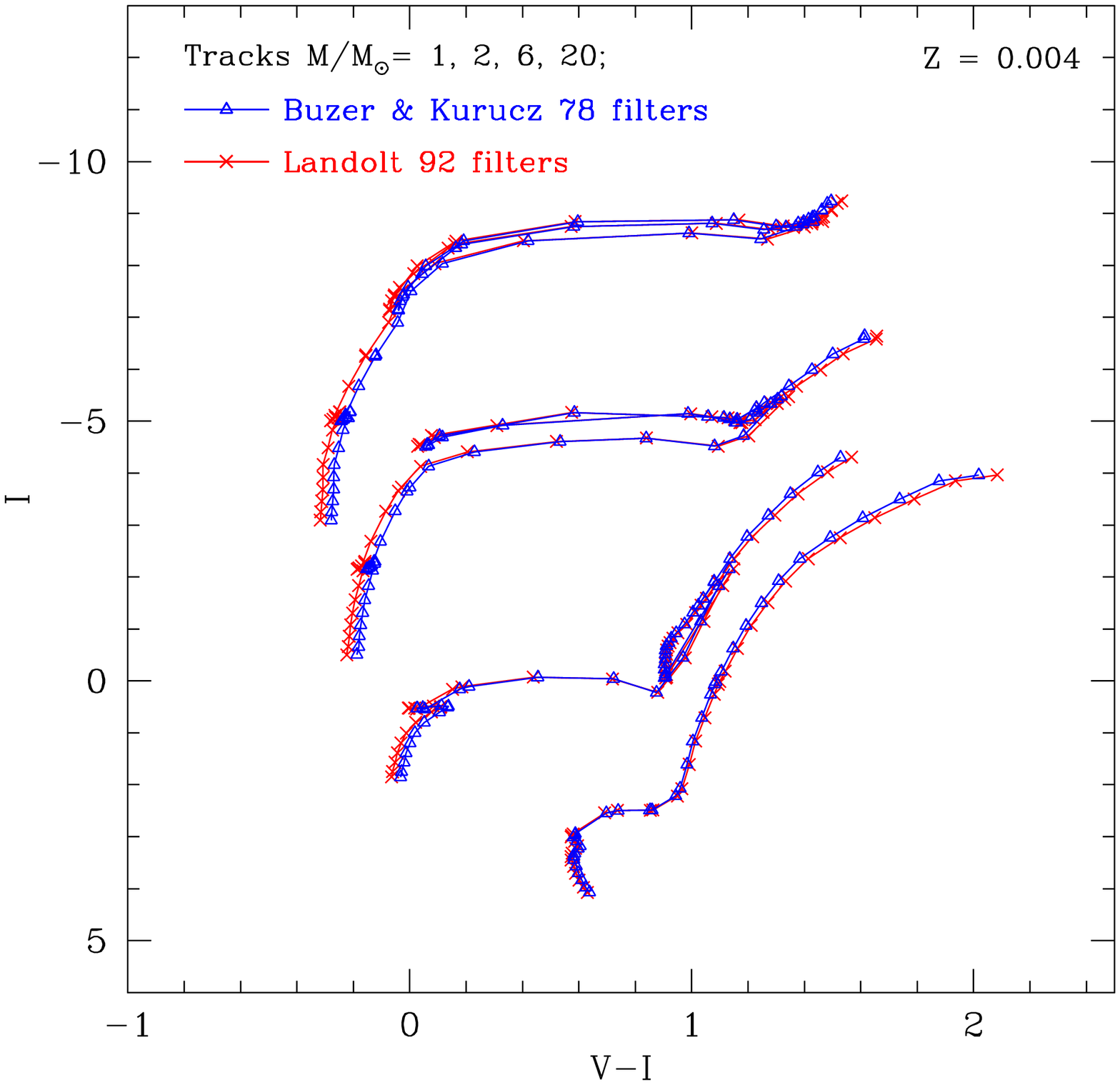}
  \caption[]
	{Left: Fagotto et al. (\cite{fagotto+94}) tracks for $Z=0.004$, 
	transformed into the $V$ $vs.$ $U-V$ 
	plane by applying bolometric corrections obtained by convolving 
	spectral energy distributions of Kurucz stellar atmospheres 
	(see text) with the two sets of $UBV$ filters: Landolt 
	(\cite{landolt92}; red lines with crosses) and 
	Buser \& Kurucz (\cite{bk78}; blue lines with triangles)
	The masses of the tracks are indicated.
	Right: The same as on the left, 
	but for $I$ vs. $V-I$. The $V$ filter
	transmission curves are 
	from Landolt (\cite{landolt92}; red lines with crosses) and 
	Buser \& Kurucz (\cite{bk78}; blue lines with triangles). The $I$
	passband is the same in both cases.}
  \label{isotrackcompare}
\end{figure*}

Bolometric corrections (BCs) are needed to convert the theoretical 
$\log({\rm L}) - \log({\rm T}_{\rm eff})$ point into magnitude and color, to
be compared with the data.
An accurate knowledge of the photometric passbands is as important to
synthetic photometry as the knowledge of the standard stars and the proper
calibration procedure is to observational photometry. The calibration of
our data was described in detail in Rejkuba et al.~(\cite{rejkuba+01}).
We have derived a full 
set of calibration equations including the color term for the
optical photometry, while only the zero point was derived for the near-IR
photometry. Thus our photometric system consists of Landolt (\cite{landolt92})
$U$ and $V$ and ISAAC $K_{\rm s}$ (Cuby et al.\
\cite{ISAACman}\footnote{\sffamily 
http://www.eso.org/instruments/isaac/index.html}).

\begin{figure}
\centering
\includegraphics[width=7cm,angle=270]{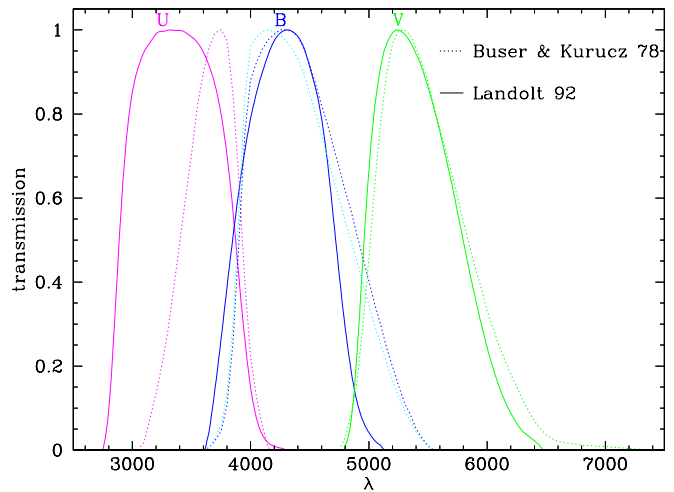}
   \caption[]
	{Comparison of the Landolt (\cite{landolt92}; full lines) 
	and Buser \& Kurucz (\cite{bk78}; dotted lines) $UBV$ passbands.}
   \label{L92_bk78filters}
\end{figure}

The Kurucz (\cite{kurucz93}) database of ATLAS9 synthetic
spectra covers a large range in 
temperatures ($3500 \leq {\rm T}_{\rm eff} \leq 50,000$~K), 
and gravities ($0 \leq \log g \leq 5$). Recently these models have been 
re-calculated by Castelli et al. (\cite{castelli+97}) and carefully compared
to stellar data. Of the two sets in Castelli's grid, the one without 
overshooting is found to yield the best match to the 
observations (Bessell et al. \cite{bcp98}).
Therefore we adopt the  ``NOVER'' files at {\sffamily
http://kurucz.harvard.edu/} for metallicities 
${\rm [M/H]} = -2.0, -1.5, -1.0, -0.5, 0.0 \quad\mbox{and} +0.5$, 
and microturbulent
velocity $\xi=2$ km~s$^{-1}$ to derive BCs
as function of gravity, effective temperature, and metallicity. This is done
by convolving the spectral energy distributions from the 
Kurucz stellar library
with filter transmission curves $S_\lambda^i$:
\begin{equation}
\langle F_i \rangle = \frac{\int\limits_0^\infty F_\lambda^i S_\lambda^i\,
d\lambda}{\int\limits_0^\infty S_\lambda^i\, d\lambda}
\end{equation}
The magnitude and the bolometric correction are then 
by definition:
\begin{eqnarray}
M_i = -2.5 \log \langle F_i \rangle + ZP_i\\
BC_i = M_{bol} - M_i
     = - 2.5 \log (\sigma T_{eff}^4) + 2.5 \log \langle F_i \rangle - ZP_i\\
\mbox{where}\quad \sigma T_{eff}^4 = \int\limits_0^\infty F_\lambda \,
d\lambda
\end{eqnarray}
The calibration of the zero points in each band follows from the definition
of the $V$-band bolometric correction of the Sun to be 
${\mathrm BC}^\odot_V = -0.07$, and the colors of Vega ${\mathrm M}^{Vega}_V - 
{\mathrm M}^{Vega}_i = 0$. The spectral energy distributions
of the Sun and Vega were taken from the Kurucz library as well.

\begin{figure}
\centering
\includegraphics[width=9cm,angle=0]{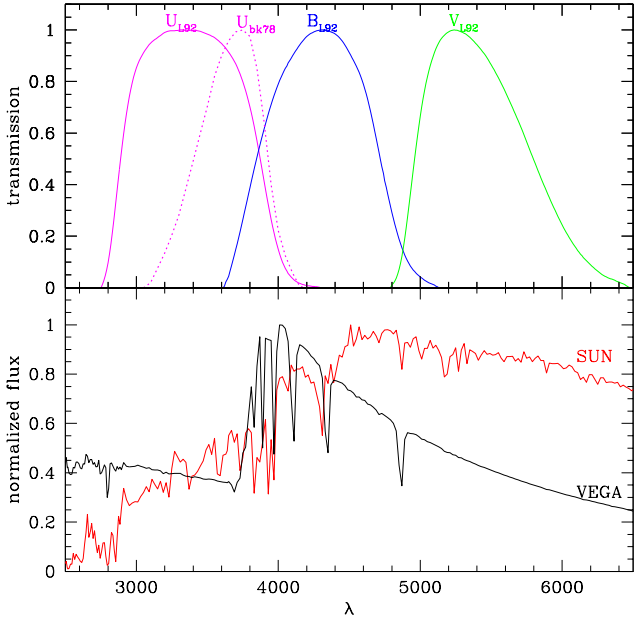}
   \caption[]
	{The spectral energy distributions of the Sun 
	  and Vega are compared with
	the Landolt (\cite{landolt92}) $UBV$ passbands (full lines) 
	and Buser \& Kurucz (\cite{bk78}) $U$ filter (dotted line).}
   \label{sun_vega}
\end{figure}

%\begin{figure}
%\centering
%\includegraphics[width=9cm]{f4.ps}
%   \caption[]
%	{The Fagotto et al. (\cite{fagotto+94}) tracks (lines with crosses and
%	triangles) for $Z=0.004$, transformed into $V$ vs. $U-V$
%        plane are compared with the Bertelli et al. (\cite{bertelli+94})
%	isochrones for $Z=0.004$ (thick lines). 
%	The tracks were transformed by convolving 
%        spectral energy distributions of Kurucz stellar atmospheres
%        (see text) with Landolt (\cite{landolt92}) $U$ and $V$ filters 
%	(red lines
%	with crosses). Overplotted are also the tracks transformed 
%	using Buser \& Kurucz (\cite{bk78}) $UB_2B_3V$ passbands where
%	($U-V$) color is defined without zero points 
%	(Eqn.~\ref{bk78nozpt}; blue lines with triangles). 
%	The $\log$(age) for isochrones 
%	and the masses of the tracks are indicated.
%	}
%   \label{isotrackbk78uv}
%\end{figure}

It is known that the Kurucz (\cite{kurucz93}) spectral energy distributions
are not very accurate for cool stars due to inadequate molecular opacities,
non-LTE, and sphericity related effects (e.g.\ Morossi et
al.~\cite{morossi+93}).  To extend our bolometric corrections to low
temperatures (${\rm T}_{\rm eff} \la 4200 {\rm K}$), we used the empirically
determined temperatures and bolometric corrections from Montegriffo et al.
(\cite{montegriffo+98}). Montegriffo et al. do not have fully developed
metallicity classes, but rather define metal-poor (${\rm [M/H]}\leq -1.0$
dex) and metal-rich (${\rm [M/H]}\geq
-1.0$ dex) samples. The differences in bolometric corrections between the
two groups span a wide range 0.05-0.4 mag/dex. We also notice that their
bolometric corrections are determined only for
giant stars. However, the last issue is not a problem, since the low
temperature stars in our case are giants with photometry in the $V$ and
$K_s$-bands. In the $UV$ only bright young stars are observed,  
the giants are too faint in the $U$-band.

We calculate the $\log g$ of each simulated star with 
($\log {\rm M}/{\rm M}_\odot, \log {\rm L}, {\rm T}_{\rm eff}$) 
using the following two equations:
\begin{eqnarray}
R^2 = \frac{L}{4 \pi \sigma T_{\rm eff}^4} \\
\log (g/g_\odot) = \log (M/M_\odot) - 2 \log (R/R_\odot)
\end{eqnarray} 
where $\log g_\odot = 4.437, \sigma = 5.67 10^{-5} {\rm erg}/{\rm cm}^2/{\rm
K}^4, {\rm L}_\odot = 3.826 10^{33} {\rm erg}/{\rm s}\, \mbox{and}\, 
{\rm R}_\odot = 6.9599 10^{10} {\rm cm}$. The linearly interpolated
bolometric correction, as a function of $\log {\rm T}_{\rm eff}$ and $\log g$, 
is then applied to
obtain magnitudes for the simulated stars.

In Fig.~\ref{isotrackcompare} we compare the evolutionary
tracks of Fagotto et al.~(\cite{fagotto+94}) for $Z=0.004$ transformed 
using 
%the procedure described above 
two different sets of $UBV$ filters.
%for $Z=0.004$ 
%of the same group (Bertelli et al.~\cite{bertelli+94}). 
In the right panel the comparison of 
$I$ $vs.$ $V-I$ shows an excellent agreement of
the colors of tracks %and isochrones
for both sets of filters, but in the left panel it is obvious
that our bolometric corrections calculated using the Landolt filters
produce wider range of colors for the
tracks in $U-V$. We investigated in detail the origin of this difference,
finding that it is entirely due to different $U$ filter used 
in calculating the bolometric corrections.

The Padova group uses Buser \& Kurucz (\cite{bk78}) $UBV$ transmission curves,
%{\bf (Bertelli et al.~\cite{bertelli+94})},
while our photometry has been calibrated to the Landolt (\cite{landolt92}) 
system. The comparison of the passbands of the two
photometric systems is given in Fig.~\ref{L92_bk78filters}. Note the large
difference in the $U$-passband wavelength coverage. 
The $U$-band is known to be the
most difficult to reproduce due to the presence of the atmospheric cut-off,
the presence of the Balmer break within that band, and the strong blanketing 
by metals (see Fig.~\ref{sun_vega}). The differences
between the $B$ and $V$-band filters are much smaller. Buser \& Kurucz
(\cite{bk78}) $UBV$ system has two $B$ filters, $B_2$ and $B_3$. Both of these
systems claim to be based on the Johnson \& Morgan (\cite{johnson&morgan})
standard system with Buser \& Kurucz  $U-B$ and $B-V$ colors defined 
as follows:
\begin{eqnarray}
(U-B)_{syn} = (U-B_2) - 1.093\\
(B-V)_{syn} = (B_3 - V) + 0.710
\end{eqnarray}

We have calculated the bolometric corrections by convolving the Kurucz stellar
library with the $UB_2B_3V$ filters of Buser \& Kurucz (\cite{bk78}).
There is an excellent agreement between the transformed tracks and Padova 
isochrones (Bertelli et al.~\cite{bertelli+94})
%(Fig.~\ref{isotrackbk78uv}) 
if the following definition of the $U-V$ color is
adopted:
\begin{equation}
(U-V)_{syn} = (U-B_2) + (B_3 - V)
\label{bk78nozpt}
\end{equation}
The difference in the zero point is due to different calibration of the zero
point bolometric corrections adopted by Buser \& Kurucz with respect to
the Padova isochrones (Girardi et al. \cite{girardi+02}). The Padova group
adopted the standard calibration of the synthetic photometry which is based
on the colors and magnitude of Vega, similar to our procedure. 

This example shows how important it is to know precisely the transmission
curves and the characteristics of the photometric system used. The
calibration of the synthetic magnitude system has to be done in the same way
as the data, by convolving  the spectral energy distributions from the stellar
library with the same filters used for the real observations.

Our data have been calibrated on Landolt's stellar catalogue 
(Landolt \cite{landolt92}), but observed
with the FORS1 telescope+CCD+filter system, so that our system's
throughput in the U band is indeed much closer to the Bessel U-filter
(with a passband similar to the Buser \& Kurucz \cite{bk78}), mainly due to the
CCD response curve. The best approach to calibrate our photometry 
would have been
to use a standard system which matches the one used in the observations
(Bessell \cite{bessell95}). 
Unfortunately, our list of observed standards has no
star in common with Bessell's (\cite{bessell95}) list.

The $U$-magnitude difference between the two systems
depends on the star's spectral energy distribution, and therefore on
temperature, gravity and metallicity. To a first approximation, the
calibration equation corrects for this systematics as a function of
the $(U-V)$ color. 
However, a residual systematic error exists, which stems from
applying to all of the observed stars the calibration
regression valid for the standard stars. In order to estimate the
residual error we have calculated the $U$-magnitude difference in the two
$U$ passbands of the complete grid of Kurucz models.
The (FORS1 - Landolt) difference of the U band Bolometric Corrections
increases from $\sim -0.2$ to $\sim +0.4$, going from the hottest to the
coolest models in the grid. Once the color equation applied to the data
is taken into account, the residual difference ranges from  $\sim$ -0.06 mag
for the blue stars $(U-V<0)$, to $\sim 0.25$ mag at $(U-V)>2$. We notice that
this is the maximum residual error that we expect in our photometry, as
it includes a very wide range of stellar parameters.
Nevertheless, our photometry will tend to be too faint for the blue
stars and too bright for the red stars.
We will discuss in the last section the impact of this unceratinty on our
results.

%
%----------------------------------------------------------------------
%
\subsection{Observational biases}

The synthetic CMD needs to be ``corrected'' for observational biases before
comparing it with the observed CMD. There are two main types of observational
biases:
\begin{itemize}
\item {\it photometric errors}: the magnitudes of stars are measured with 
	some uncertainty
\item {\it incompleteness}: some stars are lost due to crowding or high 
	background
\end{itemize}
Blending is an additional bias that is important in crowded fields, and that
combines properties of photometric errors and incompleteness. 

We used the extensive artificial stars experiments (the full description can
be found in Rejkuba et al.~\cite{rejkuba+01}) to assign the
observational biases. In the artificial star experiments the input magnitude
is given, and the output magnitude measured at each position of the input
magnitude. If the star was lost in the background noise, or if it 
was completely
blended with a much brighter object producing a magnitude difference between
the input and the measured magnitude of more than 0.75, we considered it to be
non-detected. Observational biases are applied to magnitudes to which the
appropriate distance modulus and extinction were added. 
Thus, when applying the observational biases we pick a random
object from the input table of the 
artificial star experiments that has magnitude
and color similar to the simulated star. The simulated star then receives a
realistic photometric error 
(magnitude difference between the recovered and input
magnitude in the artificial star experiment), and incompleteness (the star in
the artificial star experiment was detected or not). 

%
%----------------------------------------------------------------------
%
\section{Young stellar population observed in the $UV$ CMD}

\begin{table}
\centering
  \caption[]
	{Fraction of stars in different parts of observed CMD -- all stars, 
	blue  ($U-V<0$), intermediate color
	($0<U-V<1.5$), and red stars ($U-V>1.5$) -- 
	that survive the error selection at maximum error of 0.1, 0.2, and 0.3,
	with respect to selection with $\sigma<0.5$. Only the
	stars with $U$-band completeness limits larger than 50\% 
	are selected, corresponding to $V$-band magnitudes 
	smaller than  25.5 in the bluest part of the CMD}.
   \label{sig_error}	
	\begin{tabular}{ccccc}
\hline\hline
$\sigma$ cut & all stars & blue & intermediate & red \\
\hline
$\sigma<0.3$& 81 \% & 97\% & 70\% & 75\% \\
$\sigma<0.2$& 66 \% & 87\% & 53\% & 55\% \\
$\sigma<0.1$& 37 \% & 52\% & 29\% & 26\% \\
\hline
\end{tabular}
\end{table}

\begin{figure}
\centering
\includegraphics[width=7cm,angle=270]{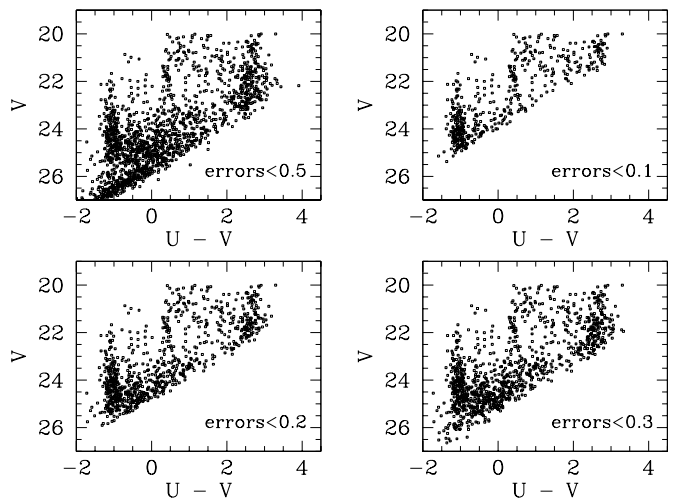}
   \caption[]
	{Color-magnitude diagrams of the young stellar populations in the
	shell field of NGC~5128 with different cuts applied on the size of the
	photometric errors in both filters $U$ and $V$.}
   \label{UVCMDs_sigcut}
\end{figure}

The observations of the young stars in the north-eastern shell field of 
NGC~5128 consist
of pairs of U and V-band images taken at ESO Paranal observatory UT1 
Very Large Telescope with FORS1 instrument (Szeifert~\cite{FORSman}\footnote{
{\sffamily http://www.eso.org/instruments/fors/userman/}})
The photometry is 50\% 
complete for magnitudes brighter than $V=27.5$ and $U=25$ (Rejkuba et al.\ 
\cite{rejkuba+01}). 
The shallower $U$-band photometry drives the incompleteness 
and the photometric errors in the
$UV$ CMD (Fig.~\ref{UVCMDs_sigcut}) and thus limits our analysis to 
magnitudes brighter than $V \sim 25.5$. 

In Fig.~\ref{UVCMDs_sigcut} we display the $V$ vs. $U-V$ CMDs of this field 
with different values of DAOPHOT errors
in both $U$ and $V$ applied as a selection criterion. 
The vertical blue sequence at 
$U-V \sim -1$ is populated with massive MS stars in NGC~5128,
while most of the red stars belong to the foreground Galactic population.
The diagonal cut-off at faint magnitudes is a consequence of the $U$-band
incompleteness and the sharp bright magnitude cut-off at $V=20$ is due to
the saturation limit of the $V$-band images.
The CMD with 
all stars that had $\sigma<0.5$ is presented in the upper
left panel. Comparing it with the other panels it is evident that many
of the red stars have large photometric errors. This is clearer
in Tab.~\ref{sig_error}, where we list fraction of stars that remain
in the CMD when a cut is applied at $\sigma$ of 0.3, 0.2 and 0.1.
Even when only the stars with the smallest 
photometric errors are selected, more than 50\% of the blue stars survive 
and the young blue sequence remains well populated. We shall use
this feature to investigate the recent star formation history in this field.

\begin{figure*}
\centering
\includegraphics[width=14cm,angle=270]{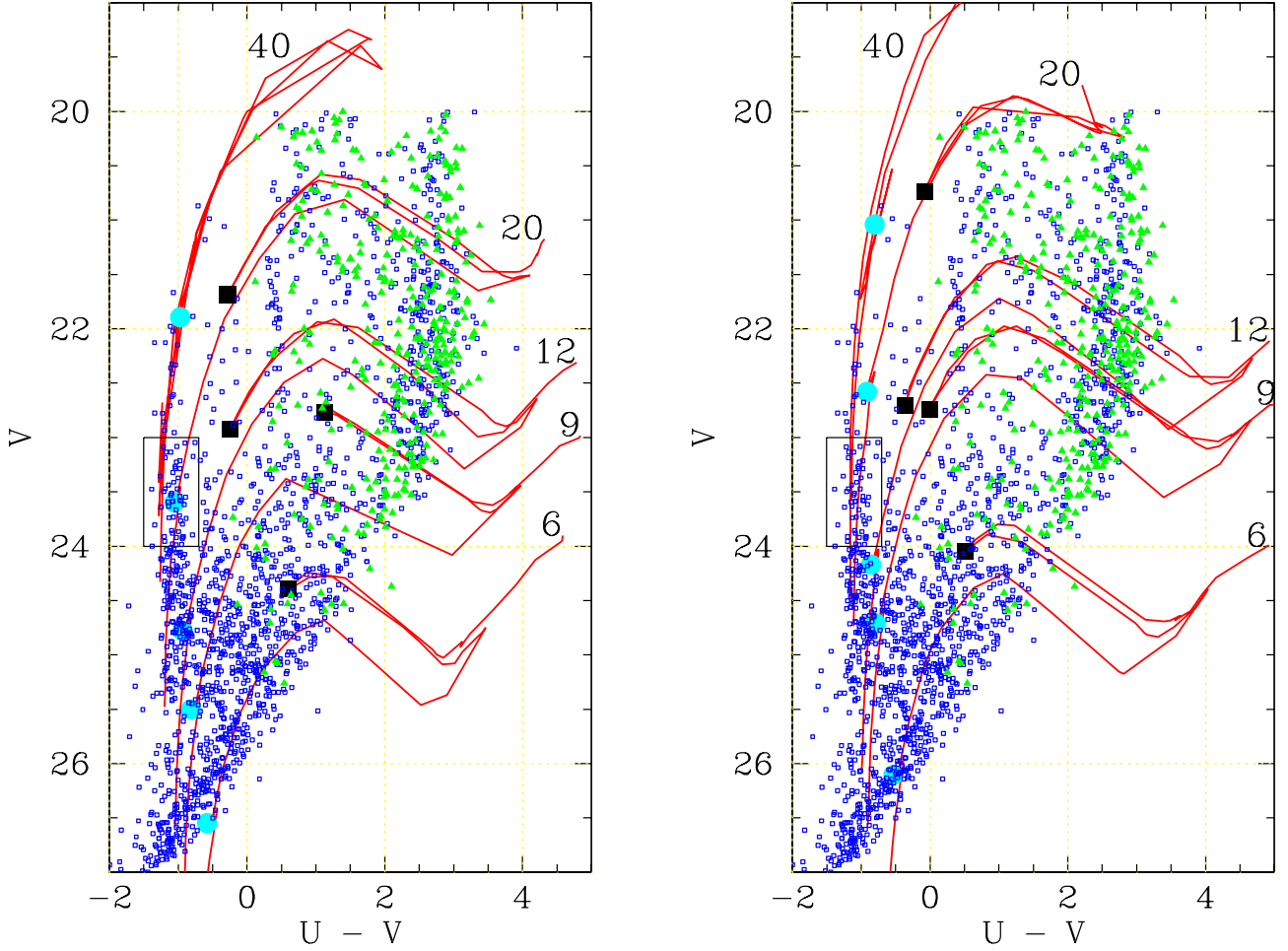}
   \caption[]
	{Comparison of the the $UV$ CMD with stellar evolutionary tracks for
	$Z=0.004$ (left) and $Z=0.008$ (right panel) and masses of 6, 9, 12,
	20 and 40 M$_\odot$. Overplotted are also the expected foreground
	stars (green filled triangles) from the Besan\c{c}on simulations 
	(see Rejkuba et al.\ \cite{rejkuba+01}). 
	Along the line connecting
	MS turn-offs (large cyan dots) and along the line 
	connecting blue edges (large black squares)
	most of the stars are expected according to the models. A box
	on the main sequence between magnitudes 23 and 24 indicates the
	area used to constrain the simulations.}
   \label{UVCMDs_tracks}
\end{figure*}

\begin{table}
\centering
  \caption[]
	{Main sequence and total life-times of the stars with masses larger
	than 4 M$_\odot$ and metallicities $Z=0.004$ and $Z=0.008$ (Fagotto et
	al.~\cite{fagotto+94}).}
   \label{life_times}	
	\begin{tabular}{rrrrr}
\hline\hline
\multicolumn{1}{c}{M(M$_\odot$)} & 
\multicolumn{1}{c}{$\tau_{MS}$(Myr)} & \multicolumn{1}{c}{$\tau_{tot}$(Myr)} &
\multicolumn{1}{c}{$\tau_{MS}$(Myr)}&\multicolumn{1}{c}{$\tau_{tot}$(Myr)} \\
              & \multicolumn{2}{c}{Z=0.004}&\multicolumn{2}{c}{Z=0.008}  \\
\hline
5.0 & 101.89 &  112.09 & 104.91 &116.05\\
6.0 & 70.98  &  76.74  & 71.63  &77.71 \\
7.0 & 52.58  &  56.39  & 52.40  &56.26 \\
9.0 & 33.07  &  35.25  & 32.42  &34.52 \\
12.0 & 20.20 &  21.48  & 19.85  &21.08 \\
15.0 & 14.35 &  15.27  & 14.03  &14.93 \\
20.0 & 9.805 &  10.48  & 9.44   &10.10 \\
30.0 & 6.37  &  6.86   & 6.18   &6.66  \\
40.0 & 5.01  &  5.44   & 4.89   &5.31  \\
60.0 & 3.85  &  4.20   & 3.84   &4.23  \\
100.0 & 3.11 &  3.38   & 3.12   &3.44  \\
120.0 & 2.98 &  3.24   & 3.01   &3.33  \\
\hline
\end{tabular}
\end{table}

Fig.~\ref{UVCMDs_tracks} shows the observed CMD with the transformed Padova
tracks superimposed, for $Z=0.004$ (left) and $Z=0.008$ (right). No
spectroscopic estimate of the metallicity of the young stars in Cen A is
available in the literature. The two considered Z values are motivated by
the Rejkuba et al.\ (\cite{rejkuba+01}) estimate, which is based on the best
fitting isochrones. Notice, however, that due to the discussed difference
between the Landolt and the Buser U filter, the Rejkuba et al.\ estimate
needs to be reconsidered with the appropriately transformed tracks.  A
distance modulus of $27.8$~mag (Soria et al.\ \cite{soria+96}) in
combination with a color excess of $\mathrm{E}(U-V)=0.62$ yields a good fit
to the position of the MS stars in Fig.~\ref{UVCMDs_tracks}. According to
the Cardelli et al.\ (\cite{cardelli+89}) extinction curve,
$\mathrm{E}(U-V)=0.62$ implies 
$\mathrm{E}(B-V)=0.35$, which is much larger than the foreground extinction 
$\mathrm{E}(B-V)=0.11$ from Schlegel et al.\ (\cite{schlegel+98}) map, 
and implies a substantial amount of internal 
(perhaps differential) reddening. Rejkuba et al.\ 
estimated $\mathrm{E}(B-V)=0.15$, due to the different U filter.
Choosing a somewhat larger distance modulus of 27.92~mag 
(Harris et al.\ \cite{harris+99}, Rejkuba \cite{rejkuba04}) does not 
change our conclusions significantly.

The magnitudes and colors of the stars belonging to the Milky Way, that are
expected to be found in a $6\farcm8 \times 6\farcm8$ field of view of FORS1,
at $l=309.5^\circ, b=19.5^\circ$, have been simulated using Besan\c{c}on
galactic model of stellar populations (Robin \& Creze \cite{robin&creze86},
Robin et al.\ \cite{robin+96}). Realistic photometric errors and
incompleteness has been applied through crowding simulations. In total, at
least 340 stars with colors
$U-V>0$  are expected to belong to the foreground Milky Way population (see
Rejkuba et al.~\cite{rejkuba+01}). They are overplotted over the observed
CMD in Fig.~\ref{UVCMDs_tracks} as (green) filled triangles. Only a small
number of stars with $(U-V)>0$ are expected to be blue and red 
supergiants evolving along the blue loops in the NGC~5128 halo.
Unfortunately, the width of the blue loops cannot be used to constrain the
metallicity both because of the huge amount of the foreground contamination
and, because of the uncertain extension of the blue loops
(Ritossa \cite{ritossa96}).

The location of the post main sequence (PMS) stars, 
blue and red supergiants, and the extension of the loops depends critically on 
the convection criteria or the mixing efficiency, 
the metallicity, the opacities at intermediate
temperatures, the mass loss and the overshooting (Renzini et
al.~\cite{renzini+92}, Ritossa~\cite{ritossa96}). At
larger metallicities the thermal conductivity of the envelope is lower
leading to a larger thermal imbalance of the stellar envelope and 
to a faster expansion to the red. The more metal-rich models have thus
redder red giant minima.
The models that adopt efficient mixing in the semiconvective region 
during the hydrogen shell burning spend most of the PMS 
life-times as blue supergiants, while the less efficient mixing
drives the runaway expansion toward the red (Ritossa~\cite{ritossa96}). 
The convective core overshoot during the MS in 
combination with efficient mixing can, however, produce
models with wider loops (e.g. Fagotto et al.~\cite{fagotto+94}). 

The overshooting and opacity, constraining T$_{\rm eff}$, drive the
amplitude of blue loops and, through the
dependence of bolometric corrections on the effective temperature
(BC(T$_{\rm eff}$)), they produce the sharp edge in the observed
CMD. In particular the 9~M$_\odot$ model for $Z=0.004$ has
a very short loop (see Fig.~\ref{UVCMDs_tracks}). The blue excursion 
of this model is terminated at low temperatures, producing a 
sharp ``red hook'' in the simulated CMD (see below). 

The extremely short evolutionary life-time for stars between the MS turn-off and blue
edge leaves a ``hole'' in the CMD. On the other hand this area 
($-0.5\la U-V\la0$ and $V\ga23$) of the observed CMD is populated.
The possible reasons include different metallicity,
inadequate input physics (combination of
overshooting and opacity parameters), differential reddening, bolometric
corrections, photometric errors and blends of several lower mass stars.
Thus we expect not to be able to reproduce well this zone. 

The evolutionary lifetimes of the plotted tracks are listed in
Tab.~\ref{life_times}. Comparing the tracks to the data we see that the
oldest MS stars in our observed CMD are $\sim40-50$~Myr old for the
$Z=0.004$ and $\sim50-65$~Myr for the $Z=0.008$ models. The faintest evolved
stars are somewhat older ($\sim 80$~Myr), but buried in the foreground
contamination. We can thus derive the most recent star formation, and check
the IMF for stellar masses larger than 7 M$_\odot$ from the bluest stars in
the CMD.

\begin{figure*}
\centering
\includegraphics[width=3.3cm,angle=270]{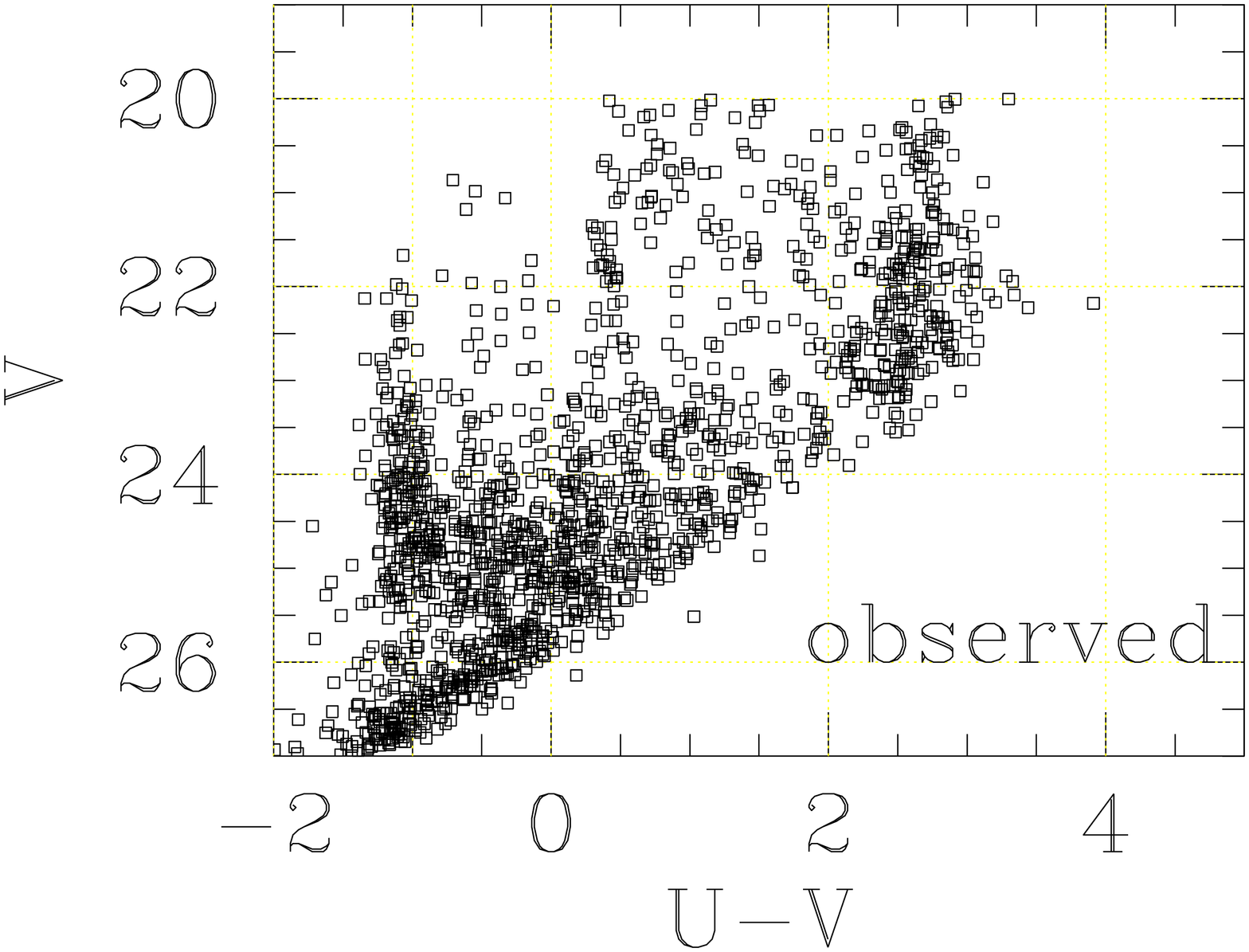}
\includegraphics[width=7cm,angle=270]{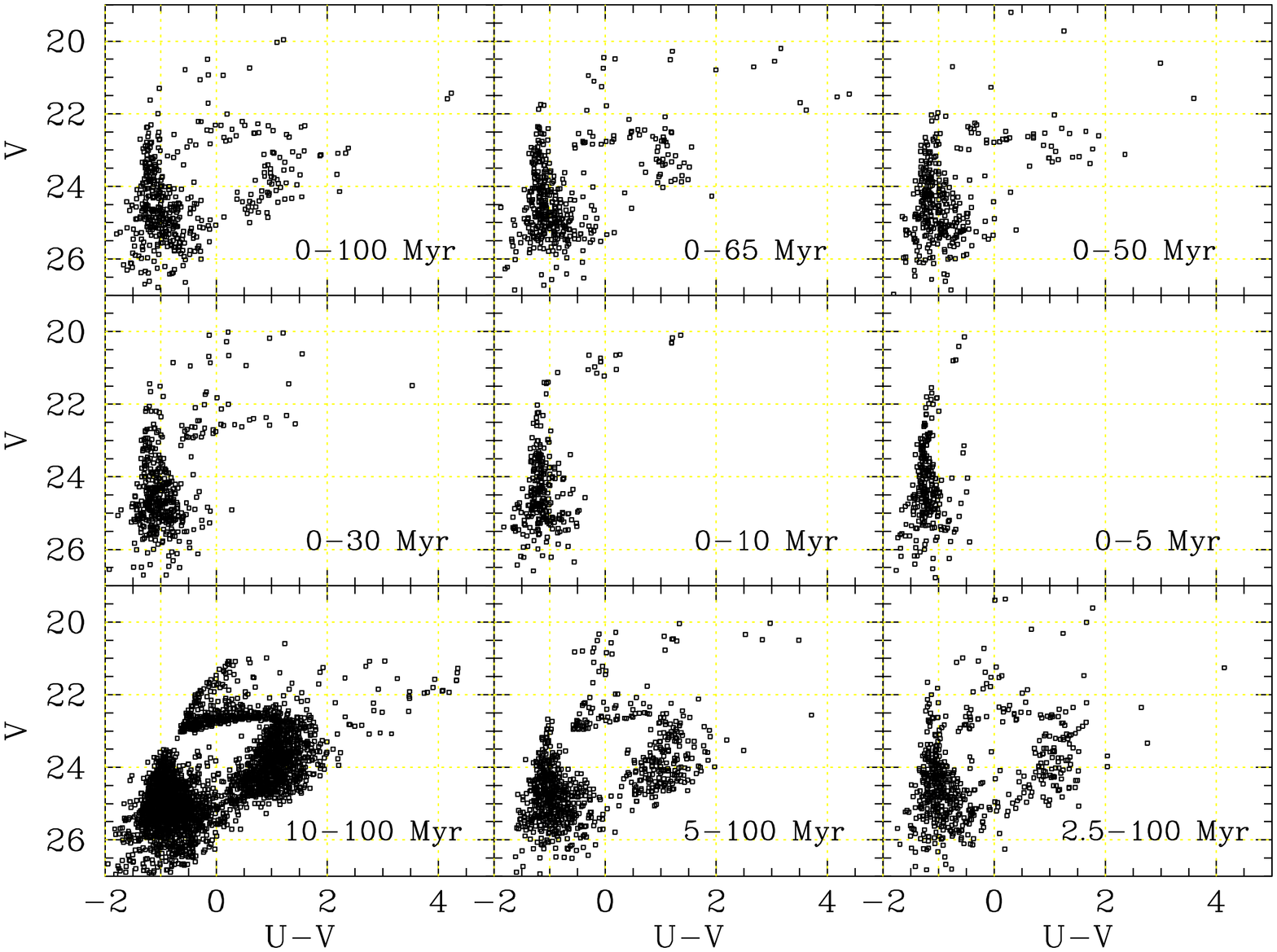}\\
\includegraphics[width=7cm,angle=270]{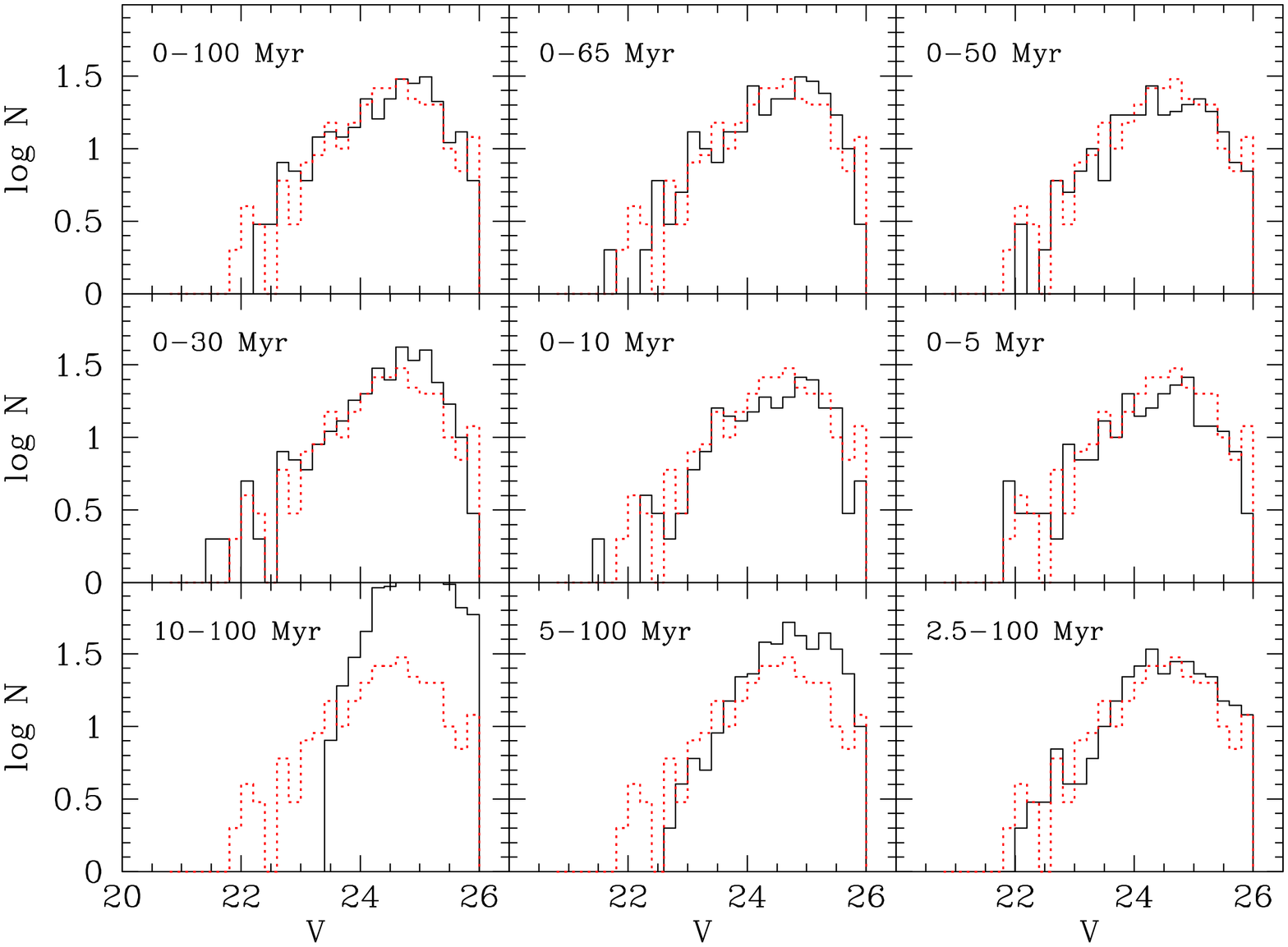}
   \caption[]
	{Synthetic CMDs for $Z=0.004$ using the Salpeter IMF ($\alpha=2.35$), 
	the appropriate distance, and a constant SFR in the interval 
	that is indicated in each panel (beginning-end
	of the burst). For comparison the observed CMD is also displayed.
	In the lower part simulated luminosity functions (solid lines) are
	compared with the observed luminosity functions (dotted lines). 
	In the  
	observed and simulated luminosity functions the color and magnitude 
	cuts are applied, and only stars bluer than 
	$(U-V)=-0.7$ and brighter
	than $V=26$ are compared.}
   \label{sim9004salp}
\end{figure*}

\begin{figure*}
\centering
\includegraphics[width=3.3cm,angle=270]{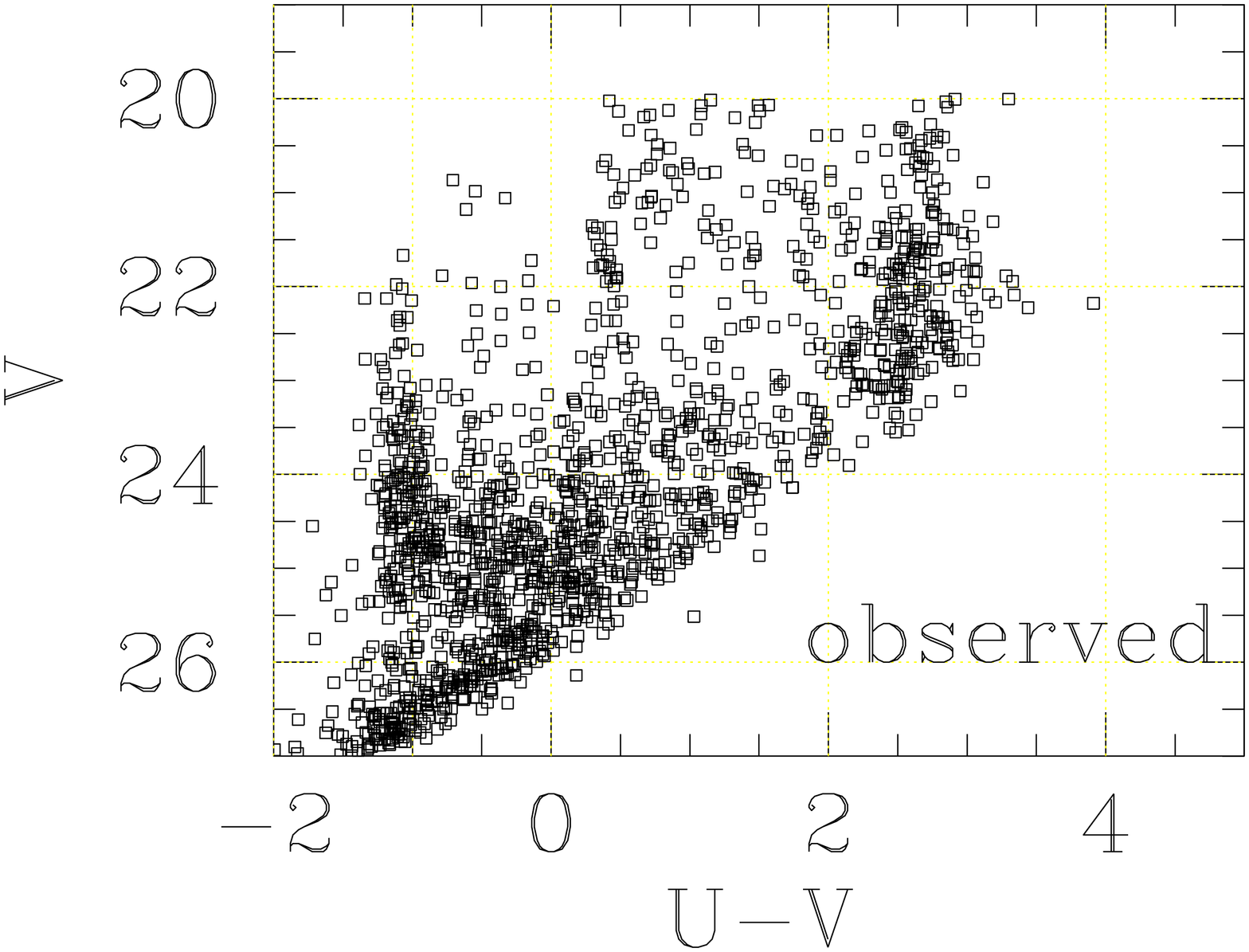}
\includegraphics[width=7cm,angle=270]{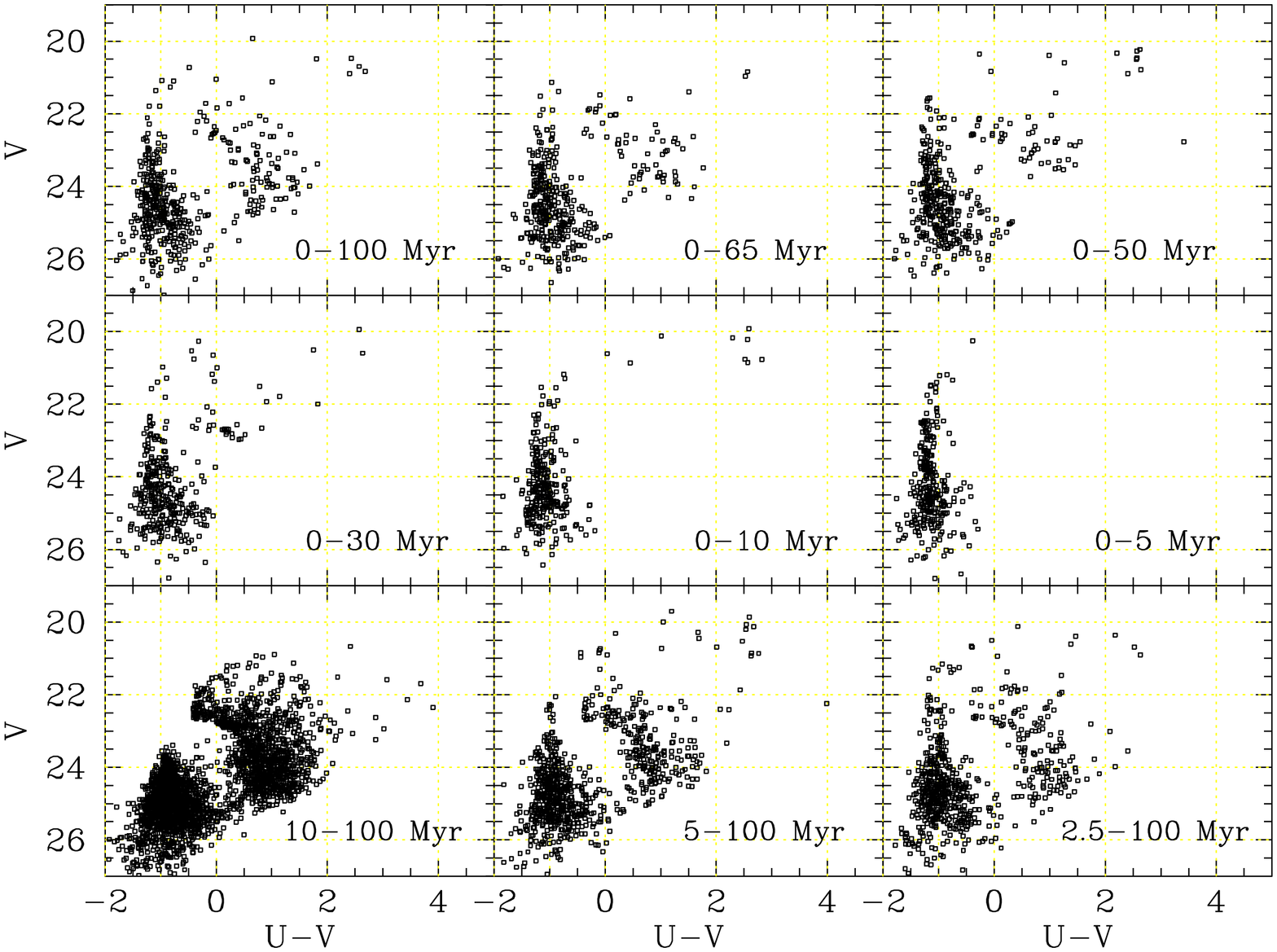}\\
\includegraphics[width=7cm,angle=270]{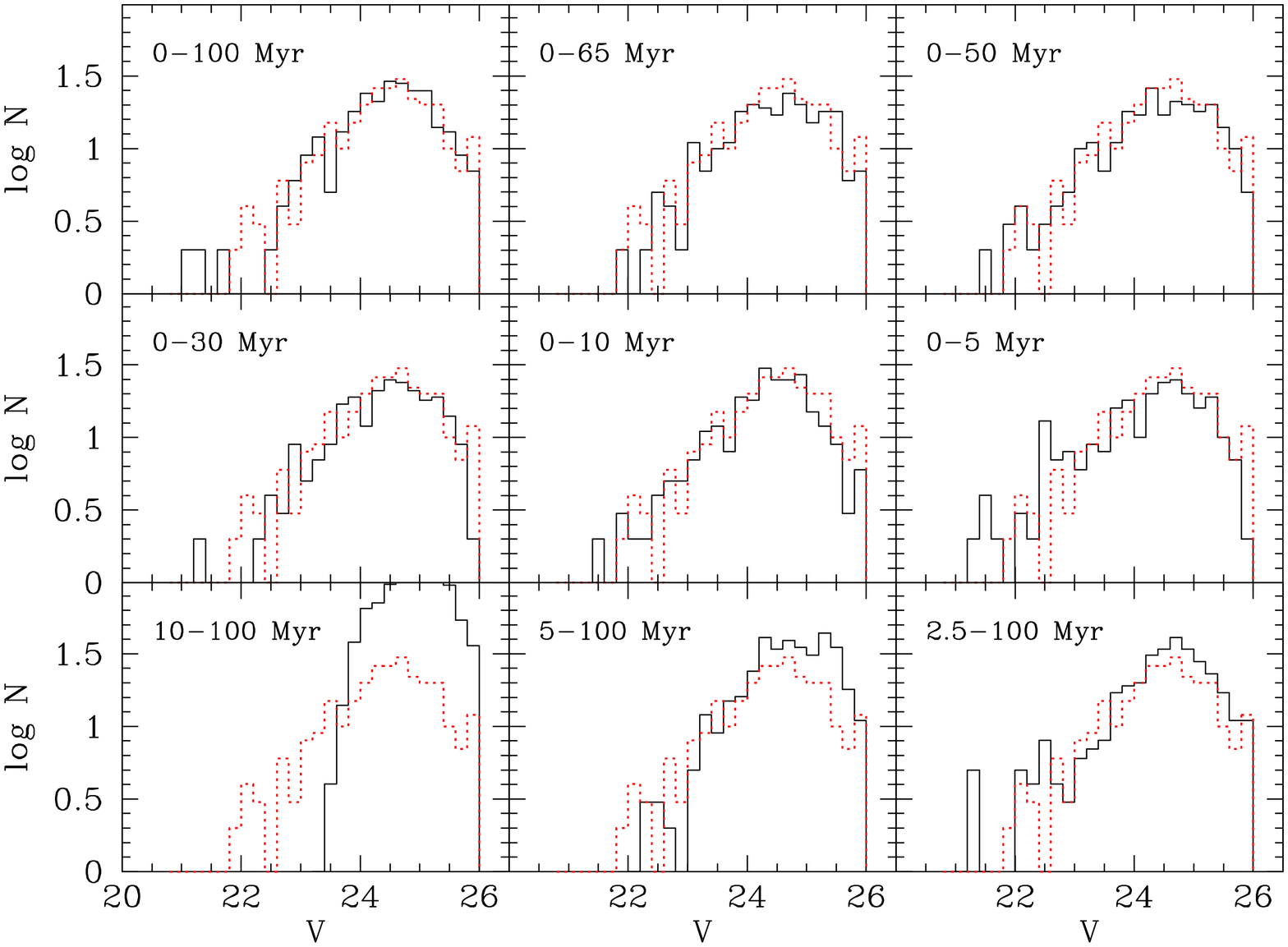}
   \caption[]
	{Synthetic CMDs for $Z=0.008$ using the Salpeter IMF ($\alpha=2.35$), 
	the appropriate distance, and a constant SFR in the interval 
	that is indicated in each panel (beginning-end of the burst).
	For comparison the observed CMD is also displayed on the left.
	In the lower panels the simulated luminosity functions (solid lines) 
	are compared 
	with the observed luminosity functions (dotted lines). In the  
	observed and simulated luminosity functions the color and magnitude 
	cuts are applied, and only stars bluer than
	$(U-V)=-0.7$  and brighter
	than $V=26$ are compared.}
   \label{sim9008salp}
\end{figure*}

Our simulations are constructed under the requirement of reproducing the 
stellar counts in a box on the MS, which contains 57 stars with 
$23 \leq V \leq 24$ and $U-V<-0.7$ (Fig.~\ref{UVCMDs_tracks}).  
The faintest stars in this box are 15~Myr old. Therefore, any SFH with the
last star formation episode ceasing earlier than 15~Myr ago would not populate
this box, and we do not consider these models. Given the short age range sampled
by the blue stars, and given their relatively smooth distribution, we
consider only episodes with constant SFR, and vary the starting ($\tau_{max}$) and 
ending ($\tau_{min}$) epochs of active SF.
We compare the synthetic and observed CMDs directly as
well as the luminosity function along the MS for the stars
brighter than $V=26$, which are not affected critically by photometric
errors and incompleteness. In the next sections we address the following
questions: ``How old are the oldest and the youngest stars in the observed
$UV$ CMD?'' and ``What is the slope of the initial mass function?'' 

%
%----------------------------------------------------------------------
%

\subsection{Age of the young stellar population in the NGC~5128 halo field}

\begin{table*}
\centering
\caption[]
	{Input parameters and results of the 
	simulated star formation histories of young stellar populations. 
	Columns 1 to 5 are identifier of the model, 
	slope of the IMF, start and end of the star 
	formation burst in Myr, respectively. In the next 8 columns 
	extracted mass,  mean star formation
	rate in the interval $\Delta\tau$, 
	the total number (N$_{\rm tot}$) and the 
	number of blue plume stars (N$_{blue}$)
	in the simulated CMDs are given for Z=0.004 and
	Z=0.008}
\begin{tabular}{ccrrrrrrrrrr}
\hline
      &      &          &                      &
\multicolumn{4}{c}{Z=0.004}&\multicolumn{4}{c}{Z=0.008}\\
\hline
model &  IMF & \multicolumn{1}{c}{$\tau_{max}$} & \multicolumn{1}{c}{$\tau_{min}$} 
&\multicolumn{1}{c}{mass}&\multicolumn{1}{c}{$\langle\mbox{SFR}\rangle$} 
& N$_{\rm tot}$ &N$_{blue}$ &\multicolumn{1}{c}{mass}& 
\multicolumn{1}{c}{$\langle\mbox{SFR}\rangle$}& N$_{\rm tot}$ &N$_{blue}$ \\
   ID &$\alpha$&\multicolumn{1}{c}{$10^6$~yr} &\multicolumn{1}{c}{$10^6$~yr} 
&\multicolumn{1}{c}{$10^6$~M$_\odot$}&
\multicolumn{1}{c}{$\mbox{M}_\odot / \mbox{yr}$} & & &
\multicolumn{1}{c}{$10^6$~M$_\odot$}&
\multicolumn{1}{c}{$\mbox{M}_\odot / \mbox{yr}$} & &\\
\hline\hline
salp01 &  2.35 & 1000 &0   & 6.79 & 0.007 &   512 &   276 & 6.44 & 0.006 &   499  &  279 \\
salp02 &  2.35 & 500  &0   & 3.97 & 0.008 &   551 &   296 & 3.38 & 0.007 &   472  &  257 \\
salp03 &  2.35 & 200  &0   & 1.38 & 0.007 &   496 &   289 & 1.47 & 0.007 &   546  &  288 \\
salp04 &  2.35 & 100  &0   & 0.66 & 0.007 &   460 &   256 & 0.61 & 0.006 &   460  &  250 \\
salp05 &  2.35 &  65  &0   & 0.44 & 0.007 &   415 &   259 & 0.35 & 0.005 &   368  &  225 \\
salp06 &  2.35 &  50  &0   & 0.29 & 0.006 &   351 &   224 & 0.31 & 0.006 &   377  &  226 \\
salp07 &  2.35 &  30  &0   & 0.23 & 0.008 &   430 &   306 & 0.19 & 0.006 &   345  &  229 \\
salp08 &  2.35 &  10  &0   & 0.08 & 0.008 &   273 &   219 & 0.08 & 0.008 &   283  &  241 \\
salp09 &  2.35 &  5   &0   & 0.07 & 0.014 &   254 &   216 & 0.07 & 0.014 &   278  &  239 \\
salp10 &  2.35 & 100  &10  & 7.73 & 0.086 &  3080 &   887 & 6.43 & 0.071 &   2685 &  801 \\
salp11 &  2.35 & 100  &5   & 1.52 & 0.016 &   816 &   353 & 1.40 & 0.015 &   806  &  339 \\
salp12 &  2.35 & 100  &2.5 & 0.90 & 0.009 &   568 &   273 & 0.94 & 0.010 &   633  &  313 \\
scal01 &  2.60 & 1000 &0   & 12.8 & 0.013 &   591 &   288 & 9.95 & 0.010 &   478  &  257 \\
scal02 &  2.60 & 500  &0   & 6.65 & 0.013 &   641 &   291 & 5.38 & 0.011 &   548  &  261 \\
scal03 &  2.60 & 200  &0   & 2.30 & 0.011 &   610 &   301 & 2.12 & 0.011 &   526  &  254 \\
scal04 &  2.60 & 100  &0   & 1.65 & 0.017 &   719 &   337 & 1.19 & 0.012 &   600  &  296 \\
scal05 &  2.60 &  65  &0   & 0.96 & 0.015 &   619 &   349 & 0.96 & 0.015 &   704  &  372 \\
scal06 &  2.60 &  50  &0   & 0.65 & 0.013 &   512 &   289 & 0.57 & 0.011 &   492  &  296 \\
scal07 &  2.60 &  30  &0   & 0.32 & 0.011 &   406 &   267 & 0.32 & 0.011 &   418  &  267 \\
scal08 &  2.60 &  10  &0   & 0.14 & 0.014 &   253 &   206 & 0.10 & 0.010 &   191  &  160 \\
scal09 &  2.60 &  5   &0   & 0.15 & 0.030 &   291 &   257 & 0.12 & 0.025 &   257  &  217 \\
scal10 &  2.60 & 100  &10  & 12.1 & 0.134 &  3497 &   921 & 10.0 & 0.112 &   3030 &  777 \\
scal11 &  2.60 & 100  &5   & 1.95 & 0.020 &   789 &   314 & 2.09 & 0.022 &   787  &  309 \\
scal12 &  2.60 & 100  &2.5 & 2.43 & 0.025 &   984 &   442 & 1.49 & 0.015 &   669  &  289 \\
imf301 &  3.00 & 1000 &0   & 34.8 & 0.035 &   939 &   366 & 28.4 & 0.028 &   743  &  314 \\
imf302 &  3.00 & 500  &0   & 21.6 & 0.043 &  1038 &   419 & 18.4 & 0.037 &   819  &  344 \\
imf303 &  3.00 & 200  &0   & 6.93 & 0.035 &   819 &   312 & 7.54 & 0.038 &   891  &  378 \\
imf304 &  3.00 & 100  &0   & 4.41 & 0.044 &   991 &   426 & 3.62 & 0.036 &   792  &  319 \\
imf305 &  3.00 &  65  &0   & 2.51 & 0.039 &   693 &   340 & 1.82 & 0.028 &   573  &  278 \\
imf306 &  3.00 &  50  &0   & 2.63 & 0.053 &   869 &   464 & 1.87 & 0.037 &   738  &  361 \\
imf307 &  3.00 &  30  &0   & 1.29 & 0.043 &   601 &   363 & 0.95 & 0.032 &   503  &  309 \\
imf308 &  3.00 &  10  &0   & 0.49 & 0.049 &   304 &   238 & 0.55 & 0.055 &   362  &  278 \\
imf309 &  3.00 &  5   &0   & 0.50 & 0.101 &   331 &   277 & 0.39 & 0.078 &   287  &  237 \\
imf310 &  3.00 & 100  &10  & 31.8 & 0.353 &  5756 &  1333 & 19.7 & 0.219 &   3267 &  784 \\
imf311 &  3.00 & 100  &5   & 6.02 & 0.063 &  1118 &   379 & 5.39 & 0.057 &   1082 &  396 \\
imf312 &  3.00 & 100  &2.5 & 5.22 & 0.053 &  1057 &   387 & 4.54 & 0.047 &   927  &  359 \\
imf201 &  2.00 & 1000 &0   & 3.87 & 0.004 &   402 &   239 & 3.75 & 0.003 &   388  &  235 \\
imf202 &  2.00 & 500  &0   & 2.11 & 0.004 &   433 &   258 & 1.68 & 0.003 &   370  &  231 \\
imf203 &  2.00 & 200  &0   & 0.93 & 0.005 &   490 &   271 & 0.68 & 0.003 &   358  &  222 \\
imf204 &  2.00 & 100  &0   & 0.40 & 0.004 &   377 &   237 & 0.43 & 0.004 &   445  &  261 \\
imf205 &  2.00 &  65  &0   & 0.32 & 0.005 &   486 &   309 & 0.26 & 0.004 &   386  &  248 \\
imf206 &  2.00 &  50  &0   & 0.19 & 0.004 &   362 &   256 & 0.20 & 0.004 &   411  &  260 \\
imf207 &  2.00 &  30  &0   & 0.10 & 0.003 &   312 &   233 & 0.14 & 0.005 &   419  &  290 \\
imf208 &  2.00 &  10  &0   & 0.05 & 0.005 &   252 &   215 & 0.03 & 0.003 &   193  &  165 \\
imf209 &  2.00 &  5   &0   & 0.03 & 0.005 &   190 &   173 & 0.03 & 0.007 &   232  &  212 \\
imf210 &  2.00 & 100  &10  & 4.25 & 0.047 &  2127 &   699 & 3.05 & 0.034 &   1705 &  517 \\
imf211 &  2.00 & 100  &5   & 0.99 & 0.010 &   693 &   318 & 0.87 & 0.009 &   625  &  311 \\
imf212 &  2.00 & 100  &2.5 & 0.52 & 0.005 &   445 &   253 & 0.45 & 0.004 &   422  &  235 \\
\hline                                              
\end{tabular}
\label{simulations}
\end{table*}

The range of models probed is listed in Tab.~\ref{simulations}. In the first
column there is the model name, and in columns 2, 3 and 4 the input
parameters regarding the IMF and the duration of the star formation are
given. The output total extracted mass in the range of 0.6 to 100 M$_\odot$,
the mean SFR for the same mass range, and the explored interval of the star
formation ($\Delta\tau$), as well as the total number of stars above the
80\% incompleteness limit in
$V$-band ($V<25.5$), and the number of simulated blue plume stars 
($U-V<-0.7$ for $V<25.5$) are listed in columns 6, 7, 8 and 9 for $Z=0.004$,
and 10, 11, 12 and 13 for models using evolutionary tracks for metallicity
$Z=0.008$, respectively. Due to the small range of ages and to the smooth
appearance of the MS, a constant star formation was considered.
Therefore the columns 6 and 10 list:
\begin{equation}
\langle \mbox{SFR} \rangle = \frac{\mbox{M}_{star}}{\Delta \tau}
\end{equation}

Four different IMF slopes have been used in the simulations. The dependence of
the simulations on the IMF slope is discussed in detail in the next section.
First we illustrate in detail the model CMDs made with Salpeter IMF
($\alpha=2.35$), varying the age range, for the two metallicities: $Z=0.004$
(Fig.~\ref{sim9004salp}) and $Z=0.008$ (Fig.~\ref{sim9008salp}).  The
duration of the star formation episode is indicated in each panel as the
starting--ending time. An ending time of 0~Myr means that the star formation
is on-going.

The sharp blue edge of the blue loops
is due to the theoretical form of the blue loops. In particular, the
early termination of the 9~M$_\odot$ model for $Z=0.004$ produces a bend
in the simulated CMDs. This transition is much smoother for the $Z=0.008$
models. The appearance of the CMD suggests that the 9~M$_\odot$ model 
for $Z=0.004$ should have bluer blue edge. The high degree of 
contamination prevents us to draw a robust conclusion, but the 
$U-V$ vs. $V$ CMD seems particularly suited to constrain the shape 
of the loops, and could be applied in stellar systems with spectroscopic 
abundance determination.

\begin{figure*}
\centering
\includegraphics[width=13.8cm,angle=270]{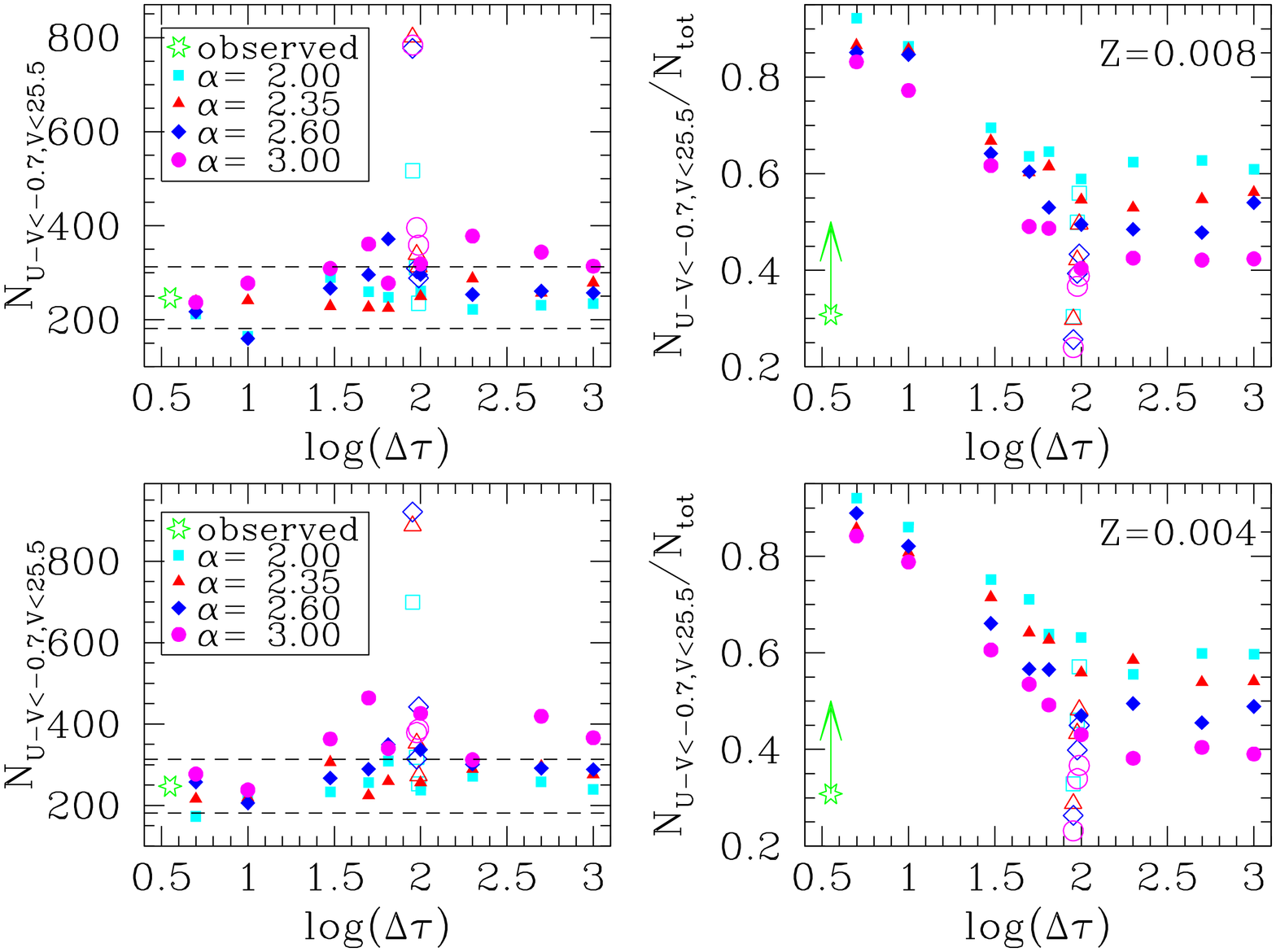}
   \caption[]
	{Left panels: Number of stars in the blue plume
	($U-V<-0.7$) for $V$ brighter than $25.5$ vs. logarithm of 
	star formation
	duration is shown for the models with $Z=0.008$
	(top) and $Z=0.004$ (bottom). The observed number of stars 
	within the same
	limits is indicated with a green star. Dotted lines
	indicate $3 \sigma$ Poissonian fluctuation limits 
	in the number of observed and simulated stars. 
	The x-axis value for the observed point is
	arbitrary. Different symbols are used for models 
	with different IMF slopes
	as indicated in the legends. The models for which 
	star formation was terminated 2.5, 5 and 10 Myr ago are plotted with
	open symbols, and the models for which the star formation is 
	still on-going
	have filled symbols. 

	Right panels: The ratio of the
	blue plume vs. total number of simulated stars. The
	observed point (corrected for the foreground contamination)
	is only a lower limit due to a large uncertainty in
	foreground contamination.}
   \label{ratios}
\end{figure*}

As the star formation lasts longer it is possible to observe the excursions
of the stars in PMS phases. The reddest parts of the CMD,
however, remain very sparsely populated due to the much shorter evolutionary
timescales in the red with respect to the blue.  Less massive 
(${\rm M}\la15{\rm M}_\odot$) red supergiants evolve less rapidly. 
Few of them are visible in the upper red part of the simulated CMDs 
in which star formation was not terminated.  
Also, as the star formation gets longer, the MS
gets populated with stars of smaller masses, the total number of
which is governed by the IMF.

In the direct comparison of the simulated and observed CMDs, and the
corresponding luminosity functions (Fig.~\ref{sim9004salp} and
\ref{sim9008salp}) it can be seen that all the simulations with on-going
star formation fit the data equivalently well irrespectively of the 
maximum stellar age ($\tau_{max}$). 
The luminosity function peak is best fitted with
models with longer $\tau_{max}$ ($\ga30-50$~Myr), 
but the differences between various models 
are within the Poissonian errors. Hence, using only MS
stars ($U-V<-0.7$) it is not possible to 
constrain $\tau_{max}$. 

On the other hand, the simulations in which the star formation has stopped
5 or 10 million years ago, have many more stars in the lower MS
and around the blue edge of the loops than observed. In order to
populate the box on the MS between magnitudes
23 and 24, which contains the youngest MS stars (younger than $14-15$~Myr) 
and thus the ones with the shortest evolutionary timescales, many more 
lower mass stars were produced in these simulations. Stopping the SF
at 2.5~Myr produces a MS luminosity function in agreement with the
observations. 

In the left panels of Fig.~\ref{ratios} we plot the number of stars in the
blue plume for all the models with $Z=0.008$ (top) and $Z=0.004$
(bottom). The observed value is shown as a green star.
Different symbols are used for models with different IMF slopes as indicated
in the legend. The dashed lines 
indicate the 3$\sigma$ Poissonian deviation from the observed 
value to which the random errors of the models 
have been added in quadrature. The models outside of the dashed 
lines in this figure are ruled out.
So, the total number of MS stars and their luminosity function 
constrain $\tau_{min}$ to be at most 2.5~Myr. The impact of the IMF on this constraint
is discussed in the next section.

Although PMS
stars cannot be used to select the best fitting model due to large
uncertainties as discussed above, in the observed CMD 
there are 17 bright ($21<V<23$) and
blue stars ($-0.7<U-V<0$) most probably in the PMS
phases. They are found in the area
of the CMD where there is no contamination by foreground Galactic stars. 
The models that have active star formation only over the last $5-10$~Myr,
have one or no stars in this area of the CMD. Thus, the star formation in this
field must have been going on for more than 10~Myr.

Better limits to the duration of the star formation can be obtained by
comparison between the number of stars in the MS and in the PMS phases. 
When comparing the observed and the simulated number of stars in the red
part of the CMD, it is necessary to subtract the expected number
of foreground stars according to the Galactic models.
The ratio of the blue plume ($U-V<-0.7$), and the total number
of stars versus the duration of the star formation 
for the models described in Tab.~\ref{simulations} is presented 
in the upper right panel of the Fig.~\ref{ratios} 
for $Z=0.008$, and lower right panel for $Z=0.004$
models. This ratio decreases with increasing $\tau_{max}$, because
the PMS phases become more populated.

The inflexion point in these diagrams corresponds to the maximum
look-back time and hence the maximum age of the burst $\tau_{\rm  
max}$. It is $\sim$100~Myr for the models
with $Z=0.008$ and $\sim90$~Myr for the $Z=0.004$ models. 
Of course, this value is
the lower limit for the duration of the star
formation. The older stars are not observed due to the incompleteness in
the $U$-band, hence the ratio of the blue vs. total number
of stars becomes constant for longer star formation durations.

The observed ratio of the blue vs. total number of stars in NGC~5128, 
which has been corrected for the expected amount of foreground contamination,
is indicated with a star in Fig.~\ref{ratios}. 
The arrow shows that it is only a lower limit 
due to the large uncertainty in contamination which affects
in particular the PMS phases. The number of stars redder than $(U-V)>0$
in the observed CMD is much larger than in all the 
simulations which reproduce correctly the number of stars
in the blue plume. 
Since this ratio is low, even after the
foreground contamination subtraction, it points, independently from the
evidence on number of MS stars, toward long $\tau_{max}$.

%
%----------------------------------------------------------------------
%
\subsection{The initial mass function}

\begin{figure*}
\centering
\includegraphics[width=13.8cm,angle=270]{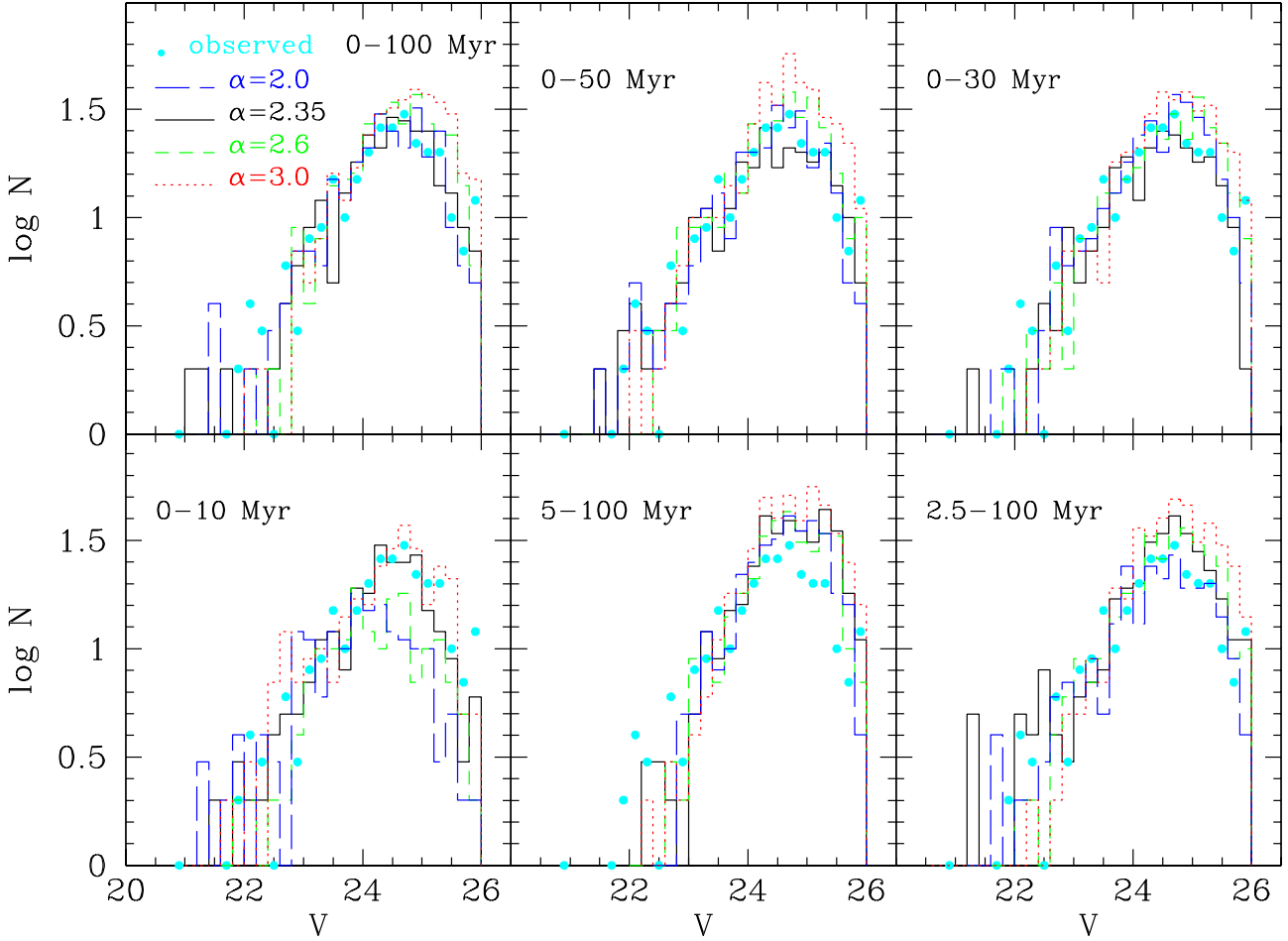}
   \caption[]
	{The observed MS luminosity function (cyan dots) is compared with
	synthetic luminosity functions for different SFHs for $Z=0.008$ and
	four different IMF slopes ($\alpha=2.0,\, 2.35,\, 2.6,\,
	\mathrm{and}\, 3.0$) as indicated in the legend.
	Synthetic luminosity functions correspond to models calculated
	adopting the appropriate distance, reddening and a 
	constant SFR in the interval 
	that is indicated in each panel (beginning-end of the burst).
	In the  
	observed and simulated luminosity functions color and magnitude 
	cuts are applied, and only stars bluer than
	$U-V=-0.7$ and brighter than $V=26$ are compared.}
   \label{IMFs008}
\end{figure*}

The observations of the young stars on the MS can be used
to constrain the initial mass function for the range of masses observed,
${\rm M}\ga5$~M$_\odot$. The comparison of the simulated and
observed luminosity functions along the MS is presented for a set
of models with on-going SF lasting for 100, 50, 30, and 10~Myr, as well as
for two models in which the SF was stopped 5 and 2.5~Myr ago.
Fig.~\ref{IMFs008} compares the observed luminosity function (dots) with the
models for Z=0.008 and four different IMF slopes: Salpeter IMF (slope
$\alpha=2.35$; solid black line), Scalo IMF (slope $\alpha=2.6$ for 
${\rm M}\leq3.5$~M$_\odot$; Scalo~\cite{scalo86}; 
green short dashed line), IMF
with slope $\alpha=2$ (blue long dashed line), and IMF with slope $\alpha=3$
(red dotted line).  

The observed slope of the MS between $22.5<V<24.5$ is
$\Delta \log N / \Delta V = 0.47$ as derived by least square fitting of the
observed luminosity function in this magnitude range. 
The results for the models with $Z=0.004$ and $Z=0.008$ are equivalent and we
plot only $Z=0.008$ models in Fig.~\ref{IMFs008}. 
The models with IMF slope $\alpha=2.35$ provide the best fit of the
luminosity function for both metallicities, $Z=0.004$ and $0.008$. 
Slightly shallower ($\alpha=2$) and steeper 
($\alpha=2.6$) slopes for IMF are also acceptable,
while the $\alpha=3$ slope produces too many stars in the lower mass bins
in all the simulations. Note that the models where the SF 
has stopped 2.5~Myr ago  
are well fitted with shallower IMFs, and they would exclude
Scalo or steeper IMF slopes. 

These results are confirmed by the $\chi^2$
tests applied to the luminosity functions of the blue stars 
and are also summarized in Fig.~\ref{ratios}. 
In the left panels the dashed lines exclude the models with IMF slope
$\alpha=3$, because the simulated number of blue plume stars is always
larger than observed (having fixed in the simulations the number of stars in 
a box on the MS). In the right panels the above mentioned saturation
value (or the inflexion point) of the number of PMS vs.\ the total number of
stars depends on the IMF slope. The steeper the IMF, the more favoured the
PMS. Even though the observed ratio would favour $\alpha=3.00$, this IMF slope is
excluded by the excess number of stars in the blue plume, which is the more robust
constraint. Therefore it seems that 
the contamination of the red part of the CMD 
is indeed much larger than that given by the Besan\c{c}on counts as already
noted by Rejkuba et al.\ (\cite{rejkuba+01}).  Additionally, there may be
few compact background galaxies contaminating the red part of the CMD.

Our simulations do not include binary stars. The presence of
binaries would increase the IMF slope by an amount that depends
on the slope and the fraction of stellar mass in binary or
multiple stellar systems. According to Sagar \& Richtler
(\cite{sagar&richtler91}) if the fraction of binaries is more than 50\% 
the IMF slope increases by 0.3 to 0.4. As the binary fraction in our field
is not known, it is possible that the actual IMF slope is somewhat steeper
than $\alpha=2.35$.

\begin{figure}
\centering
\includegraphics[width=7cm,angle=270]{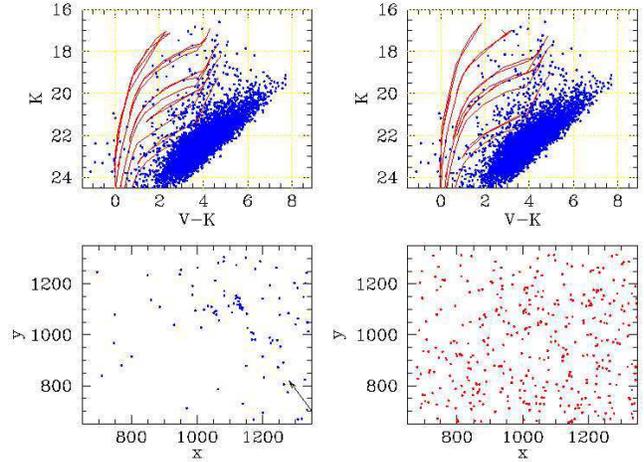}
   \caption[]
	{The comparison of the the $VK$ CMD with stellar evolutionary tracks
        for $Z=0.004$ (upper left) and $Z=0.008$ (upper right panel)
	and masses of 6, 9, 12,
        20 and 40 M$_\odot$. In the lower panels the spatial distribution
	of the blue 
	($V-K<2$ and $K<24$) stars on the left and red
	($V-K>5$ and $19<K<21$) stars on the right 
	is shown. Note the alignment of the blue stars along the
	direction of the radio jet indicated by the arrow
	in the lower right corner of the left panel.}
   \label{F1_cmdKtracks}
\end{figure}
%
%----------------------------------------------------------------------
%

\section{Young stellar population in the $VK$ CMD}

Fig.~\ref{F1_cmdKtracks} shows $VK$ optical-near IR CMD for this field 
constructed from the VLT FORS1 $V$-band and ISAAC $Ks$-band 
photometry published by Rejkuba et al.~(\cite{rejkuba+01}). 
Most of the stars are distributed along the red giant branch and 
are thus older than $\sim 1$ Gyr. There is also a large number of 
bright AGB stars, including several hundred Mira variables, located
above the tip of the RGB (Rejkuba et al.~\cite{rejkuba+03}). 
Some $\sim 110$ stars with colors $2 \la V-K \la 4$ 
are expected to belong to foreground Galactic population.
Several dozens of stars in the bluest part of the diagram ($V-K<2$) 
do not appear in the control halo field 
(see Rejkuba et al.\ \cite{rejkuba+01}).
They have very small photometric errors (90\% has $\sigma<0.2$),
and are bright in the optical. All but two stars are detected in the 
$U$-band images as well. 
We plot the spatial distribution of the blue
($V-K<2$ and $K<24$) and red ($V-K>5$ and
$19<K<21$) stars in the shell field
 in Fig.~\ref{F1_cmdKtracks}. While the reddest objects spread evenly
over the field, the majority of the blue stars are aligned in the direction
of the jet just like in the optical CMDs. 

The low number of young stars observed in the much smaller field of view
of ISAAC prevents us from drawing conclusions
from their observations. The comparison with the Padova stellar
evolutionary tracks (in the upper panels of Fig.~\ref{F1_cmdKtracks}) 
for two different metallicities $Z=0.004$ (left) and
$Z=0.008$ (right), indicates again a very small difference between the models. 

%There are only a handful of young stars in common, identified both in the $U$
%and $K$ frames. 
The $VK$ CMD gives complementary information about the older 
stellar population in the halo. However, due to large incompleteness in the red
parts of this CMD (see Rejkuba et al.~\cite{rejkuba+01}), the SFH of the old
stars in NGC~5128 will be investigated with deeper HST+ACS observations.

%
%----------------------------------------------------------------------
%

\section{Discussion and Conclusions}

The recent star formation history in the north-eastern halo field in the 
giant elliptical galaxy NGC~5128 was investigated with the means of 
synthetic CMDs. 
The comparison of the isochrones and the data has to be done
with great care, using the correct filter transmission curves to compute a
consistent set of bolometric corrections.
The $U$-band is known to be the most difficult to reproduce due to the
presence of the atmospheric cut-off, to the presence of the Balmer
break within that band, and to strong blanketing by metals. Additionally 
due to large differences between the $U$-band filters and a possible miss-match
in stellar properties ($\log \mathrm{g, } \, T_{eff}$ and [M/H]) 
between the calibrators and program objects, it is possible to introduce 
systematic errors if the calibrators do not rely on the same photometric 
system as the observations. This affects mostly our reddening estimate and
very little other results.

The new reddening estimated by matching the locus of the observed MS in
the $U-V$ plane with the models, using the corrected
bolometric corrections for $U$ passband and assuming the Galactic extinction
law (Cardelli et al.~\cite{cardelli+89}) is
$\mathrm{E}(B-V)=0.35$, significantly larger than
previously obtained $\mathrm{E}(B-V)=0.15$ (Rejkuba
et al.~\cite{rejkuba+01}). It should be noted, though, that Rejkuba et al.\ 
estimate is obtained from the comparison of mostly foreground stars 
with isochrones,
and is in fact in good agreement with $\mathrm{E}(B-V)=0.11$ from Schlegel
et al.~(\cite{schlegel+98}) maps. This implies a rather significant internal 
extinction in this field. While some internal extinction in the areas of 
recent star formation could be expected, we would like to stress that this 
result has large uncertainty due to above mentioned difficulties to reproduce 
correctly the $U$-band. The maximum possible systematic error introduced
by the calibration to the Landolt photometric system is of the order of 
0.04-0.08 in the $U$-band for the bluest stars. This translates to a possible 
over-estimate of reddening of $\mathrm{E}(B-V)=0.07-0.14$. Assuming the maximum
possible systematic error, the internal extinction in this field is reduced to
about $\mathrm{E}(B-V) \sim 0.10$.

Unfortunately, on the basis of the $U$- and $V$-band photometry 
alone it is not possible to
constrain the metallicity of the stars due
to reddening--distance--metallicity degeneracy of the models. However, the 
comparison of the synthetic and observed CMDs gave a number of 
quantitative constraints to the recent SFH in the studied halo field.

We have constructed simulated CMDs of young stars in the halo, 
having fixed the 
number of stars in a box on the bright portion of the MS, for two metallicities
Z=0.004 and Z=0.008. The parameters that were varied were $\tau_{max}$ (the
starting epoch of the SF), $\tau_{min}$ (the ending epoch of the SF), and 
$\alpha$ (the slope of the IMF).

The blue MS stars and their luminosity function 
constrain $\tau_{min}$ to be smaller
than $\sim 2.5$~Myr and exclude the IMF slopes steeper than $\alpha \ga 2.6$.
The ratio of the blue vs. total number of stars favours steeper IMF slope and
longer star formation duration ($\tau_{max} \ga 100$~Myr). 
This is a robust result and is not affected by the systematic errors in 
calibration, as the slope of the MS does not change. The exact ratio of the
blue vs. total number of stars may be affected by the systematics, but not
so much to change our conclusions.

From the simulations with 
$\tau_{\rm min}=0$ and $\tau_{\rm max}=100$~Myr, 
the total extracted mass in stars in the range of
masses $0.6\le \mbox{M} \le100$~M$_\odot$ is $\sim$0.4, 0.7, or 
$1.3\times 10^6$~M$_\odot$, for IMF slopes of 2 and 2.35, or 2.6,
respectively. The SFR associated with these IMFs and mass limits 
are approximately 0.004, 0.007, and 0.013~M$_\odot$ yr$^{-1}$, 
for the three IMF slopes in the above order, and it is independent 
of $\tau_{max}$. Moreover the smooth appearance of the MS, 
the absence of gaps and breaks in the slope of the MS luminosity 
function, indicates that most probably no major
fluctuations in the SFR have occurred over the sampled time interval.
The reported values of the SFR depend on a combination of the 
MS and PMS stars, for which the possible systematic errors in magnitudes 
(that go in different direction for these two 
groups) partially cancel out. We estimate that our results are not
changed strongly by the systematics in calibration.

At the distance of 3.6~Mpc, the observed field surveys a projected area
of approximately 45 kpc$^2$, so that the derived
star formation rates range from $\sim 9 \times 10^{-5}$ to $\sim 3
\times 10^{-4}$ M$_\odot$ yr$^{-1}$ kpc$^{-2}$. This is 
lower than the typical SFR quoted for the solar neighbourhood
($4.2 \times 10^{-3}$ M$_\odot$ yr$^{-1}$ kpc$^{-2}$; 
Timmes et al.~\cite{timmes+95}), and is on the low side
of the SFRs derived via theoretical simulations for dwarf
irregular galaxies in the Local Group (ranging from approximately $10^{-4}$
to $10^{-2}$ M$_\odot$ yr$^{-1}$ kpc$^{-2}$; Marconi
et al.~\cite{marconi+95}, Tolstoy~\cite{tolstoy96}, Gallart et
al.~\cite{gallart+96a}). For example, the star formation rate in WLM
dIrr within the last million years was $1.3 \times
10^{-3}$ M$_\odot$ yr$^{-1}$ kpc$^{-2}$ (Dolphin \cite{dolphin00}).

On a more speculative side, it is possible to estimate the
maximum duration of the future star formation.
The fuel for this recent star formation is provided by the large
HI gas cloud in the vicinity of the field (Schiminovich et
al.~\cite{schiminovich+94}). Assuming the constant SFR and a 
100\% efficiency in converting gas to stars, 
we can estimate the gas depletion timescale as:
\begin{equation}
\tau_g = \frac{\mbox{M}_{\rm HI}}{\mbox{SFR}}
\end{equation}
In the north-eastern shell very close to where the star formation
is observed, $2.1 \times 10^7$~M$_\odot$ of HI have been
detected (Schiminovich et al.~\cite{schiminovich+94}) as well as $1.7 
\times 10^7$~M$_\odot$ of H$_2$ (Charmandaris et
al.~\cite{charmandaris+00}).
At the constant star formation level of 0.006 M$_\odot$~yr$^{-1}$ 
it would take $\sim$100~Myr to use up all the molecular gas and $\sim$230~Myr 
to use all the hydrogen. 

Our results place a lower limit of $\sim 100$~Myr to the last 
merging event and 
a lower limit to the time of the formation of the jet in Cen~A. To set better
constraints on the starting epoch of the last SF episode deeper CMDs and a
more robust determination of the foreground contamination are necessary.

%
%----------------------------------------------------------------------
%

\begin{acknowledgements}
We thank D. Minniti and D. Silva for useful comments. We acknowledge an 
anonymous referee for his/her very useful comments, and especially for 
pointing out the uncertainty of the photometry
due to the calibration procedure.
\end{acknowledgements}

\appendix
\section{Random extraction of a star with a given mass and age}

We use a Monte Carlo algorithm to randomly select a mass and an age according
to a given distribution. The mass is extracted according to a single slope
(e.g. Salpeter IMF; Salpeter~\cite{salpeter55}),
\begin{equation}
dN = A M^{-\alpha} dM
\end{equation}
or double slope initial mass function (e.g. Gould IMF; Gould et
al.~\cite{gould+97}) with slope
$\alpha_1$ within a mass range $M_1\leq M \leq M_2$ and $\alpha_2$ within
$M_2 \leq M \leq M_3$. Here ``A'' is the normalisation factor 
that depends on the
upper and lower mass limits used in the models. The choice of the slope is a
free parameter. 

The age distribution is adopted to be either flat, 
representing a constant star formation rate, 
\begin{equation}
\Psi (t) = C \qquad\mbox{for}\qquad  \tau_{min} \leq t \leq \tau_{max}
\end{equation}
or exponentially decreasing, 
\begin{equation}	
\Psi (t) = C e^{-t/\tau}  \qquad\mbox{for}\qquad 
\tau_{min} \leq t \leq \tau_{max}
\end{equation}
with given limits for the age of the oldest stars ($\tau_{max}$) and for 
the duration of the star formation episode ($\Delta \tau = 
\tau_{max} - \tau_{min}$). 

We have used a single metallicity simulator to produce synthetic CMDs for the
young stellar population. This is a good approximation, because the spread
in ages of the young stars ($\mbox{few} \times 10$ Myr)
is too short for a significant metal enrichment to occur.
However in the simulation of the
optical-near IR CMDs that are dominated by the old and intermediate-age
population, evolution of the metallicity would have to be assumed. 
%Thus in these
%simulations we also extract a random variable $Z$ distributed according to a
%metal-enrichment law. The possible choices are a flat distribution between the
%minimum and maximum metallicity ($0.0001\leq Z \leq 0.04$), 
%a closed box, or a numerical distribution. 

%
%----------------------------------------------------------------------
%
%\appendix
\section{Interpolation of the stellar evolutionary models}

We have adopted two different sets of models: Padova models  
(Fagotto et al.~\cite{fagotto+94}) and models computed by Cassisi \&
Salaris (\cite{cassisi&salaris}) and extended by Bono et 
al.~(\cite{bono+97a,bono+97b}).
For the simulations of single
metallicity star formation histories used to compare 
with the youngest stars
in NGC 5128, we use the Padova models by Fagotto et al. (\cite{fagotto+94}). 
These models have tracks for stellar masses ranging from 0.6 to 120
$M_\odot$. The newer set of Padova models (Girardi et
al.~\cite{girardi+00}) are not computed for stars with masses larger than 7
$M_\odot$ such as those 
that are observed in the north-eastern halo field of NGC~5128.

Given the extracted mass and age, the luminosity and 
temperature of the star were determined via the linear interpolation 
in a set of evolutionary tracks. 
Each track consists of a number of points
(age, mass, $\log({\rm L}/{\rm L}_\odot)$, $\log({\rm T}_{\rm eff})$). 
We have divided each track into
portions according to stellar evolutionary phase, and interpolated linearly
between the adjacent tracks within the same portion. 
Typically, the portions of the tracks are:
\begin{itemize}
\item Main sequence (MS) - ending either at core H exhaustion or at the start of
the first runaway expansion
\item Subgiant branch - ending at the base of the RGB
\item RGB - ending at the tip of the RGB
\item Horizontal branch - ending either at the core He exhaustion or at the
base of the AGB
\item Early AGB - ending at the first thermal pulse
\end{itemize}
For the high mass stars (${\rm M} \geq 20\,
{\rm M}_\odot$), whose evolution is affected by mass loss, we divide the
evolution in two major phases (MS and post main sequence), and
choose conveniently the description of the post main sequence (PMS)
phase, inserting further sub-phases when needed.

In most of the cases the $\log({\rm L})$ 
and $\log({\rm T}_{\rm eff})$ 
of the extracted star
are interpolated between the values of adjacent masses read at the same
fractional age in the appropriate evolutionary phase:
\begin{equation}
t_{ph} = (t - t_0)/(t_1-t_0)
\end{equation}
where $t_0$ and $t_1$ are respectively the ages at the beginning and 
at the end of the appropriate evolutionary phase and $t$ is the age of the
extracted object. In few evolutionary phases when luminosity and/or
temperature presented a maximum or minimum at variable $t_{ph}$, we inserted
a break $t_{ph}(t_{ph,br})$ and effectively divided that phase 
into two sub-phases. Log(L) and $\log({\rm T}_{\rm eff})$ 
of the extracted 
star at the extracted age are then found by interpolation at:
\begin{eqnarray}
t_{ph} / t_{ph,br} ~\mbox{for ages}~  t_{ph} \leq
t_{ph,br}\\
(t_{ph} - t_{ph,br}) / (1-t_{ph,br}) ~\mbox{for ages}~ 
t_{ph} \geq t_{ph,br} 
\end{eqnarray}

When the interpolation has to be done also in metallicity, the 
simple linear interpolation algorithm described above is not
enough. It is necessary to perform a bilinear interpolation, because
the stars of the same masses, but different metallicities
have different ages. Thus for each extracted mass one has to first
interpolate in  $\log({\rm L})$ and $\log({\rm T}_{\rm eff})$ 
for a given age, and then again
for a given metallicity. In order to simplify the problem 
Zoccali et al. (\cite{zoccali+03}) adopted a set
of isochrones as a starting point. Using
the tabulated values of  $\log({\rm L})$ and $\log({\rm T}_{\rm eff})$ 
for stars of given mass in an isochrone the interpolation 
is done linearly within isochrones of the same age and
different metallicities. The Padova models are not adequate for these
simulations due to the very low number of points in each isochrone 
(approx. 30-55 points for each track) and it is better to use the Cassisi models
(Cassisi \& Salaris~\cite{cassisi&salaris}, Bono et
al.~\cite{bono+97a,bono+97b}) which have much more detailed evolution 
with more than 600 points per track. 
Given a mass and metallicity, the luminosity and temperature of the
star are determined through the linear interpolation in a set of 
isochrones. The interpolation in  $\log({\rm L})$ and 
$\log({\rm T}_{\rm eff})$  is made for
the mass that has the same evolutionary phase of the extracted mass:
\begin{eqnarray}
ph = m / m_1 \qquad\mbox{for objects on the main sequence}\qquad \\
ph = m_0 + (m - m_0)/(m_1-m_0) \qquad\mbox{for later phases}\qquad
\end{eqnarray}
where $m_0$ and $m_1$ are the masses at the beginning and the end of the
evolutionary phase. 
The evolutionary phases are as before: main sequence, subgiant branch, RGB,
core-helium burning and AGB. 

%

%
%____________________________________________________________________________
%

\end{document}